\documentclass[pra,twocolumn,aps,amssymb,showpacs,superscriptaddress]{revtex4-1}

\usepackage{epsfig}
\usepackage{graphicx}
\usepackage{amsmath} 
\usepackage{amssymb}
\usepackage{enumerate}
\usepackage{hyperref}
\usepackage[normalem]{ulem}
\usepackage{cancel}

\usepackage{color}
\definecolor{rosso}{rgb}{1,0,0}
\definecolor{verde}{rgb}{0,1,0}
\definecolor{blue}{rgb}{0,0,1}
\definecolor{verdescuro}{rgb}{0,0.5,0.5}
\definecolor{rossoscuro}{rgb}{0.7,0.3,0}
\definecolor{bluscuro}{rgb}{0.3,0,0.7}
\definecolor{magenta}{rgb}{1,0,1}

\begin{document}

\title{Josephson effect at finite temperature along the BCS-BEC crossover}

\author{V. Piselli}
\affiliation{School of Science and Technology, Physics Division, Universit\`{a} di Camerino, 62032 Camerino (MC), Italy}
\author{S. Simonucci}
\affiliation{School of Science and Technology, Physics Division, Universit\`{a} di Camerino, 62032 Camerino (MC), Italy}
\affiliation{INFN, Sezione di Perugia, 06123 Perugia (PG), Italy}
\author{G. Calvanese Strinati}
\email{giancarlo.strinati@unicam.it}
\affiliation{School of Science and Technology, Physics Division, Universit\`{a} di Camerino, 62032 Camerino (MC), Italy}
\affiliation{INFN, Sezione di Perugia, 06123 Perugia (PG), Italy}
\affiliation{CNR-INO, Istituto Nazionale di Ottica, Sede di Firenze, 50125 (FI), Italy}


\begin{abstract}
The Josephson current-phase characteristics, that arise when a supercurrent flows across two fermionic superfluids separated by a potential barrier, can be controlled by varying either the inter-particle coupling
or the temperature.
While the coupling dependence has been addressed in detail both theoretically and experimentally for an attractive Fermi gas undergoing the BCS-BEC crossover, a corresponding study of the temperature dependence of the Josephson characteristics is still lacking in this context.
Here, we investigate the combined coupling \emph{and\/} temperature dependence of the Josephson characteristics in a systematic way for a wide set of barriers, within ranges of height and width 
that can be experimentally explored.
Our study smoothly connects the two limiting cases, of non-overlapping composite bosons at low temperature described by the Gross-Piatevskii equation, and of strongly overlapping Cooper pairs near 
the critical temperature described by the Ginzburg-Landau equation. 
In this way, we are able to explore several interesting effects related to how the current-phase characteristics evolve along the BCS-BEC crossover as a function of temperature and of barrier shape. 
These effects include the coherence length outside the barrier and the pair penetration length inside the barrier (which is related to the proximity effect), as well as the temperature evolution of the Landau criterion 
in the limit of a vanishingly small barrier.
A comparison is also presented between the available experimental data for the critical current and our theoretical results over a wide range of couplings along the BCS-BEC crossover.
\end{abstract}

\maketitle

\section{Introduction} 
\label{sec:introduction}

The stationary Josephson effect is one of the most striking consequences of the spontaneous symmetry breaking of the phase $\varphi(\mathbf{r})$ of the complex gap (order) parameter 
$\Delta(\mathbf{r}) = |\Delta(\mathbf{r})|e^{i \varphi(\mathbf{r})}$ at a spatial point $\mathbf{r}$, which is at the essence of the macroscopic quantum coherence of superconductivity \cite{Barone-1982}.  
At a microscopic level, this phenomenon amounts to a coherent tunneling of Cooper pairs across a geometrical constraint (like a potential barrier) placed at the interface between two superconductors \cite{Josephson-1962}. 
It results in a \emph{characteristic current-phase relation} $J(\delta \phi)$ between the stationary supercurrent $J$ and the asymptotic phase difference $\delta \phi$ across the potential barrier, 
which in its simplest version has the form $J(\delta \phi) = J_{\mathrm{c}} \sin(\delta \phi)$ where $J_{\mathrm{c}}$ is the maximum (critical) value attained by the supercurrent for a given barrier.

Extending the study of the Josephson effect from the BCS limit, where it has traditionally been considered \cite{Barone-1982}, to the BEC limit, where coherence effects
based on composite bosons (dimers) emerge, casts the Josephson effect under a broader perspective \cite{SPS-2010}.
Here, we are referring to the BCS and BEC limits of the BCS-BEC crossover, respectively, whereby by tuning the inter-particle coupling (or the density) the system 
evolves from a BCS state with largely overlapping pairs of (opposite spin) fermions described by Fermi statistics, to a BEC state with dilute fermion dimers described by Bose statistics \cite{Physics-Reports-2018}.
In the process, the physical system passes through an intermediate situation (commonly referred to as the ``unitary'' regime), where the pair size is comparable with the average 
inter-particle spacing.

In principle, the Josephson characteristics $J(\delta \phi)$, together with the associated critical current $J_{\mathrm{c}}$, can be calculated in terms of the Bogoliubov-de Gennes (BdG) equations \cite{BdG-1966}, for any value of the inter-particle coupling and of the temperature, as well as for arbitrary shapes of the potential barrier.
In practice, however, this calculation was performed throughout the BCS-BEC crossover only at zero temperature \cite{SPS-2007}\cite{SPS-2010}, where a potential barrier embedded in an otherwise homogeneous superfluid extending to infinity on both sides was considered, with the fermionic coupling kept unmodified under the barrier (a situation encountered experimentally with ultra-cold Fermi gases and to which we shall refer to
as an SsS Josephson junction).
Related experiments with ultra-cold Fermi gases to determine the coupling dependence of $J_{\mathrm{c}}$ were then realized also in the (near) zero-temperature regime \cite{Ketterle-2007} \cite{Moritz-2015}.
Recently, the interest in the Josephson effect with ultra-cold Fermi gases has been revived \cite{Roati-2020}, with future plans for extending these experimental studies to finite temperature so as to approach the superfluid critical temperature $T_{c}$.
A corresponding theoretical study of the Josephson effect at finite temperature in the context of the BCS-BEC crossover thus appears appropriate and timely, in order to exploit in a systematic way the dependence of the Josephson characteristics and of related quantities on temperature, coupling, and barrier shape. 
This is the main purpose of the present paper.

As already mentioned, the BdG equations (or, alternatively, the Gor'kov equations for superconductors \cite{Gorkov-1958}) could, in principle, be utilized to address the Josephson effect at finite temperature, even across the BCS-BEC crossover. 
Solving for these equations represents, however, a formidable numerical task, mainly due to the fact that a considerable amount of information about the detailed spatial structure of the single-particle wave functions that solve the BdG equations has to be stored for a wide range of energies, before proceeding to an effective averaging procedure over these details so as to obtain the physical quantities of interest (such as the local gap parameter 
$\Delta(\mathbf{r})$). 
For these reasons, \emph{approximate} versions of the Gor'kov equations with a spatially varying gap parameter have been utilized at finite temperature, with the invariable restriction to the BCS regime where the Cooper pair size is large compared to the average inter-particle spacing. 
Various types of Josephson junctions (such as SIS, SNS, and SS'S) have been considered in this context \cite{Golubov-2004}, both in the clean limit via the Eilenberger equation \cite{Eilenberger-1968} \cite{Rammer-2007} and in the dirty limit via the Usadel equation \cite{Usadel-1959} \cite{Rammer-2007}, as well as close to the critical temperature $T_{c}$ via the Ginzburg-Landau equation \cite{Gorkov-1961}.

In this paper, consideration of the Josephson characteristics and of the associated physical quantities for various types of junctions is extended not only at finite temperature but also over an extended range of inter-particle coupling, from the BCS regime traditionally considered for the Josephson effect, up and beyond the unitary regime of the BCS-BEC crossover.
To achieve this task, we solve numerically up to self-consistency the non-linear differential LPDA equation for the spatially dependent gap parameter $\Delta(\mathbf{r})$, that was obtained in Ref.~\cite{SS-2014} by a suitable double-coarse-graining procedure on the BdG equations (where the acronym LPDA stands for Local Phase Density Approximation).
As further shown in Ref.~\cite{SS-2017}, the length scale of the ``granularity'' associated with this coarse-graining procedure corresponds to the Cooper pair size $\xi_{\mathrm{pair}}$,
for any coupling throughout the BCS-BEC crossover and temperature below $T_{c}$ \cite{PS-1994,Palestini-2014}.
This implies that the spatial profiles (of  the magnitude and phase) of $\Delta(\mathbf{r})$ obtained by the LPDA approach are expected to be reliable when they vary \emph{smoothly\/} 
over a spatial extent not smaller than the coarse-graining length scale $\xi_{\mathrm{pair}}$.
This condition is automatically satisfied by the Ginzburg-Landau (GL) equation for Cooper pairs \cite{Gorkov-1961} and by the Gross-Pitaevskii (GP) equation for composite bosons \cite{PS-2003}, which can be derived 
(either from the BdG equations \cite{Gorkov-1961,PS-2003} or from their coarse-grained LPDA version \cite{SS-2014}) in the appropriate limits, namely,
in the BCS limit at temperatures close to $T_{c}$ for the GL equation and in the BEC limit at low temperature for the GP equation.
Conversely, the results of the differential LPDA equation for $\Delta(\mathbf{r})$ are expected to be less reliable in the BCS limit at low temperature, when $\xi_{\mathrm{pair}}$ becomes comparable with the minimal extent of the spatial variations of $\Delta(\mathbf{r})$ determined by the healing length \cite{Physics-Reports-2018}.
Within these provisions, however, the advantage of solving the LPDA equation in the place of the original BdG equations stems from a considerable reduction of computation time and memory space, which makes it possible to consider physical problems for which solving for the BdG equations would be computationally too demanding.
Consideration of the Josephson effect at finite temperature along the BCS-BEC crossover falls within this class of problems.
In any case, it should be remarked that solving the LPDA equation can at most recover the same physical contents obtained by solving the BdG equations, which are the inhomogeneous version of the BCS approach to fermionic superfluidity \cite{BdG-1966}.

To simplify the numerical calculations, we shall limit to considering a rectangular potential barriers of height $V_{0}$ and width $L$. 
Among these, we will consider values of $V_{0}$ and $L$ such that the junction can be assimilated to an SIS structure with small transparency \cite{Golubov-2004}. 
This situation corresponds, in particular, to the experimental setup involving ultra-cold Fermi gases of Ref.~\cite{Roati-2020}.
A detailed study of the Josephson characteristics will also be performed for values of $V_{0}$ and $L$ that correspond to SNS and SS'S structures \cite{Golubov-2004}.

The main results obtained in this paper are as follows:  

\noindent
(i) A systematic study is performed for the spatial profiles of the magnitude and phase of the gap parameter, as functions of inter-particle coupling, temperature, and barrier width and height. 
[Only a limited choice of these profiles will be explicitly shown below.]
Complementary information will be extracted from these profiles, as they extend both outside and inside the barrier.

\noindent
(ii) The shapes of the Josephson current-phase characteristics are obtained at finite temperature along the BCS-BEC crossover for a wide set of barriers. 
For all couplings, these shapes are found consistent with the classification made in the BCS limit \cite{Golubov-2004}.
Only in few circumstances (namely, for large $V_{0}$ or when $T \simeq T_{c}$) the Josephson characteristics recover the simple sinusoidal form of a tunnel junction \cite{Josephson-1962} \cite{Barone-1982}.

\noindent
(iii) The healing length $\xi_{\mathrm{out}}$ for restoring the bulk value of the gap parameter \emph{outside} the barrier is determined along a given Josephson characteristic. 
In all cases, this length acquires its maximum in correspondence to the critical current $J_{c}$ of the given characteristic. 

\noindent
(iv) The coherence length $\xi_{\mathrm{in}}$ \emph{inside} the barrier is determined from the exponential decrease of $J_{c}$ for increasing value of the barrier width. 
Physically, $\xi_{\mathrm{in}}$ will be associated with the proximity effect \cite{Deutscher-1969}, whereby superconducting correlations penetrate from outside to the interior of a wide enough barrier 
(which, in this limit, exemplifies an SNS junction \cite{Golubov-2004}).

\noindent
(v) Consideration is given to the behaviour $J_{c} \propto (T_{c}-T)^{\eta}$ when the temperature $T$ approaches $T_{c}$ for given barrier. 
In particular, on the BCS side up to unitarity, the exponent $\eta$ is found to vary with continuity for increasing $V_{0}$ and $L$, from the value $1.5$ (large transparency) to the value $2$ (small transparency), 
in line with what is expected to occur deep in the BCS limit under the same conditions \cite{Golubov-2004}.

\noindent
(vi) Loosely speaking, the GL regime is expected to be recovered in the extreme BCS regime for temperature sufficiently close to $T_{c}$. 
Here, we endow this statement with a quantitative meaning and delimit the GL sector in the temperature-phase diagram of the attractive Fermi gas,
by determining numerically how the results of the GL approach deviate from those of the LPDA approach for physical quantities of interest in the context of the Josephson effect.

\noindent
(vii) The Landau criterion for fermionic superfluidity is implemented at finite temperature, when excitations present in the system affect the superfluid flow \cite{AGD-1975}. 
This will be achieved by extending to finite temperature the strategy adopted in Ref.~\cite{SPS-2010} in terms of the BdG equations at zero temperature, thereby determining the limiting value of $J_{c}$ for a vanishing barrier at given coupling in terms of the LPDA equation at finite temperature.

\noindent
(viii) Irrespective of coupling on the BCS side of the crossover, for a vanishingly small barrier $J_{c}$ is found to be proportional to the condensate fraction through a universal function that depends only on temperature. 

\noindent
(ix) Finally, a favourable comparison is found to emerge between the results of our theoretical calculations and the experimental data for $J_{c}$ reported in Refs.~\cite{Moritz-2015} and \cite{Roati-2020},
which cover a wide coupling range of the BCS-BEC crossover in the low-temperature regime and are obtained using weak and strong barriers, respectively.

The paper is organized as follows.
Section \ref{sec:theoretical_approach} sets up the LPDA approach in the presence of a supercurrent and a potential barrier.                                               
Section \ref{sec:Josephson-characteristics} applies the LPDA approach to obtain the Josephson characteristics under a variety of conditions.                 
Section \ref{sec:spatial-profiles-out} presents the spatial profiles of the magnitude and phase of the gap parameter associated with these characteristics.
Section \ref{sec:spatial-profiles-in} relates the gap profiles in the interior of the barrier to the appearance of the proximity effect.                                          
Section \ref{sec:GL-regime} determines how the GL regime gradually emerges from the LPDA approach in the temperature-coupling phase diagram.
Section \ref{sec:critical-current} addresses the temperature and coupling behaviour of the Josephson critical current for given barrier.                                 
Section \ref{sec:Landau-criterion} considers the Landau criterion for superfluidity at finite temperature, whereby a vanishingly small barrier acts as an impurity that probes the stability of the superfluid flow. 
Section \ref{sec:condensate-density} proposes a relationship between the critical current and the condensate density, valid on the BCS side of unitarity for a vanishingly small barrier.
Section \ref{sec:comparison-experiment} shows the comparison with available experimental results for ultra-cold Fermi gases.                                           
Section \ref{sec:conclusions} gives our conclusions.
The Appendix describes in detail the numerical procedures and strategies, that we have adopted to solve the LPDA equation in the presence of a supercurrent and of a potential barrier.
Throughout the paper, we set $\hbar =1$.

\section{Josephson effect at finite temperature within the LPDA approach} 
\label{sec:theoretical_approach}

\begin{figure}[t]
\begin{center}
\includegraphics[width=6.5cm,angle=0]{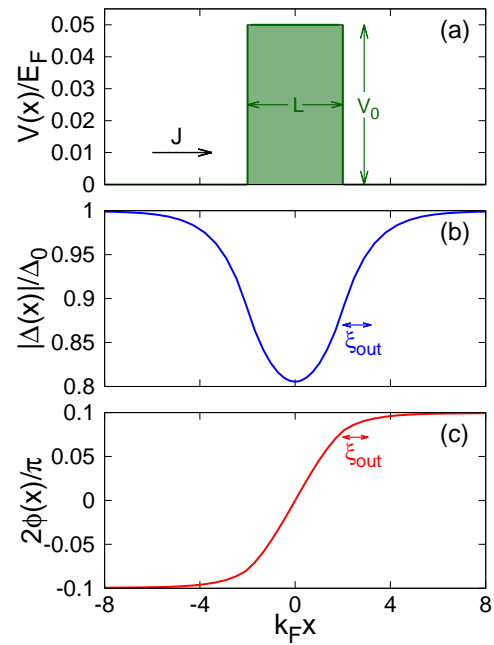}
\caption{(Color online) (a) Rectangular shape of the potential barrier of height $V_{0}$ and width $L$, on which the stationary supercurrent $J$ impinges from the left.
                                      Profiles (obtained at unitarity and for $T/T_{c}=0.5$) of the (b) magnitude and (c) phase of the gap parameter due to the presence of the barrier, which approach their bulk values $\Delta_{0}$ and $\delta \phi / 2$, respectively, 
                                      over the distance $\xi_{\mathrm{out}}$.
                                      Here, $k_{F}$ is the Fermi wave vector and $E_{F} = k_{F}^{2}/(2m)$ the corresponding Fermi energy.}                                     
\label{Figure-0}
\end{center}
\end{figure}

To deal with the stationary Josephson effect for a superfluid Fermi system that evolves along the BCS-BEC crossover at finite temperature, we adopt the geometrical set up of Ref.~\cite{SPS-2010}
and consider an SsS slab geometry, with a potential barrier $V(\mathbf{r}) = V(x)$ embedded in an otherwise homogeneous superfluid which extends to infinity on both sides of the barrier along the $x$ direction of the superfluid flow.
In addition, to simplify the numerical calculations, the slab is taken to extend to infinity along the two $y$ and $z$ orthogonal directions, and a rectangular shape centered at $x=0$ is adopted for $V(x)$  
with height $V_{0}$ and width $L$.  
The geometry of the problem with a schematic picture of the potential barrier is shown in Fig.~\ref{Figure-0}, together with the typical depression of the magnitude and increase of the phase of the gap parameter $\Delta(x)$ occurring across the barrier.

The Fermi system (with attractive inter-particle interaction and balanced spin populations) is made to span the BCS-BEC crossover by varying the dimensionless coupling parameter $(k_{F} a_{F})^{-1}$, where $k_{F}=(3 \pi^{2} n)^{1/3}$ is the Fermi wave vector associated with the bulk density $n$ and $a_{F}$ is the scattering length of the two-fermion problem \cite{Physics-Reports-2018}.
In the following, we shall mostly be interested in the coupling region $-1.5 \lesssim (k_{F} a_{F})^{-1} \lesssim +0.5$ about the unitarity limit $(k_{F} a_{F})^{-1} = 0$, where our numerical calculations based on (a coarse-grained version of) the BdG
equations \emph{at finite temperature\/} are expected to be more reliable than in BEC regime where $(k_{F} a_{F})^{-1} \gtrsim +0.5$.

\vspace{0.05cm}
\begin{center}
{\bf A. LPDA equation in the presence of a supercurrent}
\end{center}

The Josephson effect here considered requires the presence of a stationary supercurrent $J$ flowing across a potential barrier.
This condition can be dealt with in terms of the LPDA approach of Ref.~\cite{SS-2014} in the following way.

Quite generally, the LPDA equation for the spatially dependent gap parameter $\Delta(\mathbf{r})$ reads \cite{SS-2014}:
\begin{eqnarray}
- \frac{m}{4 \pi a_{F}} \, \Delta(\mathbf{r}) & = & \mathcal{I}_{0}(\mathbf{r}) \, \Delta(\mathbf{r}) + \mathcal{I}_{1}(\mathbf{r}) \, \frac{\nabla^{2}}{4m} \Delta(\mathbf{r})  
\nonumber \\                  
& - & \mathcal{I}_{1}(\mathbf{r}) \, i \, \frac{\mathbf{A}(\mathbf{r})}{m} \cdot \nabla \Delta(\mathbf{r}) \, .
\label{LPDA-equation}
\end{eqnarray}
Here, $m$ is the fermion mass,
\begin{eqnarray}
\mathcal{I}_{0}(\mathbf{r}) & = & \int \! \frac{d \mathbf{k}}{(2 \pi)^{3}} \, 
\left\{ \frac{ 1 - 2 f_{F}(E_{+}^{\mathbf{A}}(\mathbf{k}|\mathbf{r})) }{2 \, E(\mathbf{k}|\mathbf{r})} - \frac{m}{\mathbf{k}^{2}} \right\}
\label{I_0} \\
\mathcal{I}_{1}(\mathbf{r}) & = & \frac{1}{2} \, \int \! \frac{d \mathbf{k}}{(2 \pi)^{3}} 
\left\{  \frac{\xi(\mathbf{k}|\mathbf{r})}{2 \, E(\mathbf{k}|\mathbf{r})^{3}} \, \left[  1 - 2 f_{F}(E_{+}^{\mathbf{A}}(\mathbf{k}|\mathbf{r})) \right] \right.                          
\nonumber \\
& + &  \frac{\xi(\mathbf{k}|\mathbf{r})}{E(\mathbf{k}|\mathbf{r})^{2}} \, 
\frac{\partial f_{F}(E_{+}^{\mathbf{A}}(\mathbf{k}|\mathbf{r}))}{\partial E_{+}^{\mathbf{A}}(\mathbf{k}|\mathbf{r})} 
\nonumber \\
& - & \left. \frac{\mathbf{k}\cdot\mathbf{A}(\mathbf{r})}{\mathbf{A}(\mathbf{r})^{2}} \, \frac{1}{E(\mathbf{k}|\mathbf{r})} \, 
\frac{\partial f_{F}(E_{+}^{\mathbf{A}}(\mathbf{k}|\mathbf{r}))}{\partial E_{+}^{\mathbf{A}}(\mathbf{k}|\mathbf{r})} \right\} 
\label{I_1}
\end{eqnarray}
are highly non-linear functionals of the solution $\Delta(\mathbf{r})$ of Eq.~(\ref{LPDA-equation}), with
$E_{+}^{\mathbf{A}}(\mathbf{k}|\mathbf{r}) = E(\mathbf{k}|\mathbf{r}) - \frac{\mathbf{k} \cdot \mathbf{A}(\mathbf{r})}{m}$,
$E(\mathbf{k}|\mathbf{r}) = \sqrt{\xi(\mathbf{k}|\mathbf{r})^{2} + |\Delta(\mathbf{r})|^{2}}$, and
$\xi(\mathbf{k}|\mathbf{r}) = \frac{\mathbf{k}^{2}}{2m} - \bar{\mu}(\mathbf{r})$, where $\bar{\mu}(\mathbf{r}) = \mu - V(\mathbf{r}) - \mathbf{A}(\mathbf{r})^{2}/(2m)$
contains the thermodynamic chemical potential $\mu$ and the external potential $V(\mathbf{r})$.
Here and in the following, the notations $\xi(\mathbf{k}|\mathbf{r})$, $E(\mathbf{k}|\mathbf{r})$, and $E_{+}^{\mathbf{A}}(\mathbf{k}|\mathbf{r})$ signify that these three functions of the wave vector $\mathbf{k}$ (which is integrated over in the expressions 
(\ref{I_0}) and (\ref{I_1})) also depend parametrically on the spatial position $\mathbf{r}$ via the local values of $V(\mathbf{r})$, $\mathbf{A}(\mathbf{r}$), and $\Delta(\mathbf{r})$.
The temperature $T$ enters the Fermi function $f_{F}(E) = (e^{E/(k_{B}T)} + 1)^{-1}$ where $k_{B}$ is the Boltzmann constant.
Equation (\ref{LPDA-equation}) has to be solved until self-consistency is achieved.

In the above expressions, $\mathbf{A}(\mathbf{r})$ plays formally the role of an ``effective'' vector potential, although for a neutral Fermi gas its physical meaning differs 
from the vector potential of classical electrodynamics.
For instance, $\mathbf{A}(\mathbf{r}) = m \mathbf{\Omega} \wedge \mathbf{r}$ for a neutral Fermi gas embedded in a rotating trap with angular velocity $\mathbf{\Omega}$, whereby complex vortex patterns can be generated \cite{SPS-2015} (in this case, $\bar{\mu}(\mathbf{r})$ does not contain the term $\propto \mathbf{A}(\mathbf{r})^{2}$).
For the Josephson effect of interest here, when a supercurrent of magnitude $J$ impinges on the barrier from the left, $\mathbf{A}(\mathbf{r})$ in Eqs.~(\ref{LPDA-equation})-(\ref{I_1}) is identified with a constant wave vector $- \mathbf{Q}_{0}$, such that at zero temperature $J = Q_{0} n/m$ where $Q_{0} = | \mathbf{Q}_{0}|$
(in this case, the term $\propto \mathbf{A}(\mathbf{r})^{2} = \mathbf{Q}_{0}^{2}$ is present in the expression of $\bar{\mu}(\mathbf{r})$).
At the same time, the complex order parameter solution of the LPDA equation (\ref{LPDA-equation}) with $\mathbf{A}(\mathbf{r}) = - \mathbf{Q}_{0}$ reads 
$\Delta(\mathbf{r}) = |\Delta(\mathbf{r})|e^{i 2 \phi(\mathbf{r})}$, where the \emph{additional\/} phase $2 \phi(\mathbf{r})$ arises from the presence of the barrier such that 
$\varphi(\mathbf{r}) = 2 \mathbf{Q}_{0} \cdot \mathbf{r} + 2 \phi(\mathbf{r})$ is the total phase in the frame where the barrier is at rest.
The LPDA equation can be either solved by fixing the asymptotic phase difference 
\begin{equation}
\delta \phi = 2 [\phi(+ \infty) - \phi(- \infty)]
\label{phase-difference}
\end{equation}
accumulated across the barrier and determining $\mathbf{Q}_{0}$ consistently, or viceversa.

\vspace{0.05cm}
\begin{center}
{\bf B. Supercurrent flow at finite temperature} 
\end{center}

The identification of $J$ with $Q_{0} n/m$ was also made in Ref.~\cite{SPS-2010}, when dealing with the Josephson effect at zero temperature in terms of the BdG equations.
At finite temperature when thermal excitations arise, however, the superfluid current no longer coincides with $Q_{0} n/m$.
Consistently with the LPDA approach of Ref.~\cite{SS-2014}, we take the (coarse-grained) local current of the form:
\begin{eqnarray}
\mathbf{j}(\mathbf{r}) & = & \frac{1}{m} \! \left(\nabla \phi(\mathbf{r}) + \mathbf{Q}_{0} \right) \, n(\mathbf{r})
\nonumber \\
& + & 2 \int \! \frac{d\mathbf{k}}{(2 \pi)^{3}} \, \frac{\mathbf{k}}{m} \, f_{F} \! \left( E^{\mathbf{Q}_{0}}_{+}(\mathbf{k}|\mathbf{r}) \right) \, .
\label{two-fluid-current}
\end{eqnarray}
Here,
\begin{equation}
n(\mathbf{r}) =  \int \! \frac{d\mathbf{k}}{(2 \pi)^{3}} \left\{ 1 - \frac{\xi^{\mathbf{Q}_{0}}(\mathbf{k}|\mathbf{r})}
{E^{\mathbf{Q}_{0}}(\mathbf{k}|\mathbf{r})} 
\left[ 1 - 2 f_{F}(E^{\mathbf{Q}_{0}}_{+}(\mathbf{k}|\mathbf{r})) \right] \right\}
\label{local-density}
\end{equation}
is the corresponding local number density where
\begin{eqnarray}
E^{\mathbf{Q}_{0}}_{+}(\mathbf{k}|\mathbf{r}) & = & E^{\mathbf{Q}_{0}}(\mathbf{k}|\mathbf{r}) 
+ \frac{\mathbf{k}}{m} \cdot \left( \nabla \phi(\mathbf{r}) + \mathbf{Q}_{0} \right) 
\label{E+Q_0} \\
E^{\mathbf{Q}_{0}}(\mathbf{k}|\mathbf{r}) & = & \sqrt{\xi^{\mathbf{Q}_{0}}(\mathbf{k}|\mathbf{r})^{2} + |\Delta(\mathbf{r})|^{2}}  
\label{E-Q_0} \\
\xi^{\mathbf{Q_{0}}}(\mathbf{k}|\mathbf{r}) & = & \frac{\mathbf{k}^{2}}{2m} - \mu(\mathbf{r}) + \frac{1}{2m} \left( \nabla \phi(\mathbf{r}) + \mathbf{Q}_{0} \right)^{2}
\label{xi-Q_0} 
\end{eqnarray}
with $\mu(\mathbf{r}) = \mu - V(\mathbf{r})$.
The expression (\ref{two-fluid-current}) has the typical form of a two-fluid model at finite temperature \cite{Pethick-Smith-2008}, with the normal velocity vanishing for the problem of interest 
here when the normal component is at rest.
Accordingly, the expression (\ref{two-fluid-current}) for the supercurrent can also be cast in the form
\begin{equation}
\mathbf{j}(\mathbf{r}) = \frac{1}{m} \! \left(\nabla \phi(\mathbf{r}) + \mathbf{Q}_{0} \right) \, n_{s}(\mathbf{r}) \, ,
\label{current-superfluid-density}
\end{equation}
where $n_{s}(\mathbf{r})$ stands for the \emph{local superfluid density\/} which depends itself on $\mathbf{Q}_{0}$ \cite{footnote-1}.
Sufficiently far away from the barrier when $V(\mathbf{r}) = 0$, $\nabla \phi(\mathbf{r}) \rightarrow 0$ and $|\Delta(\mathbf{r})| \rightarrow \Delta_{0}$ ($\Delta_{0}$ being the bulk value of the gap parameter).
In addition, the thermodynamic chemical potential $\mu$ in the presence of the supercurrent that enters Eq.~(\ref{xi-Q_0}) equals the value of the corresponding chemical potential
in the absence of the supercurrent (for the given coupling and temperature), added by the kinetic term $\mathbf{Q}_{0}^{2}/(2m)$.
These conditions also define the bulk value $J$ of the supercurrent.

In particular, in the zero-temperature limit $J$ given by the expression (\ref{two-fluid-current}) coincides with $Q_{0} n/m$, provided the second term on the right-hand side of Eq.~(\ref{two-fluid-current}) vanishes.
This occurs insofar as the argument 
$\sqrt{ \left( \frac{\mathbf{k}^{2}}{2m} - \mu + \frac{\mathbf{Q}_{0}^{2}}{2m} \right)^{2} + \Delta_{0}^{2} } \, + \frac{\mathbf{k}}{m} \cdot \mathbf{Q}_{0}$
of the Fermi function remains positive for all $\mathbf{k}$.
Violation of this condition corresponds to the Landau criterion for the collapse of superfluidity \cite{AGD-1975}, when the relevant quasi-particles consist of pair-breaking excitations 
\cite{SPS-2010}.
At finite temperature, the second term on the right-hand side of Eq.~(\ref{two-fluid-current}) contributes to the current $J$ by decreasing the values of the bulk superfluid density $n_{\mathrm{s}}$
with respect to the bulk number density $n$.
This contribution modifies the conditions for the Landau criterion, as discussed in Sec.~\ref{sec:Landau-criterion}.

It is worth mentioning that, in the BEC limit  $(k_{F} a_{F})^{-1} \gg 1$ at low temperature, when the GP equation for composite bosons \cite{PS-2003} is recovered from the LPDA equation \cite{SS-2014}, only the first term on the right-hand side of Eq.(\ref{two-fluid-current}) contributes to $\mathbf{j}(\mathbf{r})$ and one obtains to the lowest order in $|\Delta(\mathbf{r})|^{2}$:
\begin{equation}
\mathbf{j}^{\mathrm{GP}}(\mathbf{r}) \simeq \frac{1}{m} \! \left(\nabla \phi(\mathbf{r}) + \mathbf{Q}_{0} \right) 2 \, |\Phi^{\mathrm{GP}}(\mathbf{r})|^{2}
\label{BEC-superfluid-current}
\end{equation}
\noindent
where $\Phi^{\mathrm{GP}}(\mathbf{r}) = \Delta(\mathbf{r}) \sqrt{m^{2} a_{F} / 8 \pi}$ is the GP wave function of composite bosons in this limit \cite{PS-2003}.
In the opposite BCS limit $(k_{F} a_{F})^{-1} \ll  -1$ for temperatures close to $T_{c}$, on the other hand, when the GL equation for Cooper pairs \cite{Gorkov-1961} is also recovered
from the LPDA equation \cite{SS-2014}, both terms on the right-hand side of Eq.(\ref{two-fluid-current}) contributes to $\mathbf{j}(\mathbf{r})$ and one obtains to the lowest order in 
$|\Delta(\mathbf{r})|^{2}$:
\begin{equation}
\mathbf{j}^{\mathrm{GL}}(\mathbf{r}) \simeq \frac{1}{m} \! \left(\nabla \phi(\mathbf{r}) + \mathbf{Q}_{0} \right) 2 \, |\Psi^{\mathrm{GL}}(\mathbf{r})|^{2}
\label{BCS-superfluid-current}
\end{equation}
\noindent
where $\Psi^{\mathrm{GL}}(\mathbf{r}) = \Delta(\mathbf{r}) \sqrt{7 \, \zeta(3) \, n / 8 \pi^{2} (k_{B} T_{c})^{2}}$ is the wave function of Cooper pairs in this limit \cite{Gorkov-1961,FW-1971}.
In the following, the results (\ref{BEC-superfluid-current}) and (\ref{BCS-superfluid-current}) will also be verified numerically in the respective limits.

Finally, we note that the expression (\ref{two-fluid-current}) of the local current plays also a practical role for the numerical solution of the LPDA equation in the context of the Josephson effect.
This is because, one expects on physical grounds the local profiles of $|\Delta(\mathbf{r})|$ and $\phi(\mathbf{r})$ to adapt themselves to the presence of the potential barrier 
$V(\mathbf{r})$ for given coupling and temperature, in such a way that the local condition $|\mathbf{j}(\mathbf{r})\!| = J$ should be satisfied for all $\mathbf{r}$, where $J$ corresponds to
the asymptotic value of the current far from the barrier.
In Ref.~\cite{SPS-2010}, while solving numerically for the Josephson problem throughout the BCS-BEC crossover at zero temperature with the local current $\mathbf{j}(\mathbf{r})$ expressed in terms of the eigen-solutions of the BdG equations, it was found it convenient to replace the imaginary part of the gap equation therein with the requirement of the local current being everywhere uniform.
Here, we adopt a similar strategy and replace the imaginary part of the LPDA equation (\ref{LPDA-equation}) by the condition 
\begin{equation}
|\mathbf{j}(\mathbf{r})\!| - J = 0
\label{local-condition}
\end{equation}
according to the procedure specified in detail in Appendix A, 
where $\mathbf{j}(\mathbf{r})$ is given by Eq.~(\ref{two-fluid-current}) at any temperature below $T_{c}$ for given coupling.
In practice, we have numerically verified that local deviations of the ratio $|\mathbf{j}(\mathbf{r})\!|/J$ from unity, which occur about the barrier, never exceed the value $10^{-8}$.
 
\vspace{0.05cm}
\begin{center}
{\bf C. Universal fitting function for \\ the Josephson characteristics}
\end{center}

As mentioned in the Introduction, from the way it was obtained in Ref.~\cite{SS-2014} through a double-coarse-graining procedure of the BdG equations, the LPDA 
approach is expected to yield reliable results when the spatial variations of the magnitude and phase of the gap parameter are sufficiently smooth.
In the present context of the Josephson effect with a one-dimensional geometry, the magnitude $|\Delta(x)|$ and phase $\phi(x)$ are expected to vary over a length scale 
not smaller than the (coupling and temperature dependent) healing length $\xi$ of the homogeneous system \cite{PS-1996} \cite{Palestini-2014} in the absence of the barrier.
The presence of a barrier of width $L$ makes this length scale to increase to a value $\mathcal{L}(\xi,L)$ that depends both on $\xi$ and $L$, as it will be discussed in detail in Sec.~\ref{sec:spatial-profiles-out}-C.
In particular, for given barrier and inter-particle coupling, $\mathcal{L}(\xi,L)$ increases for increasing temperature, in such a way that a characteristic phase difference $\delta\phi$ 
of order $\pi$ occurs over a larger and larger length for increasing temperature.
As a consequence, the LPDA approach is expected to perform better for increasing temperature.
This is especially true on the BCS side of unitarity where, quite generally, any local (differential) approach to determine the spatial profile of the gap parameter is expected not to be appropriate at low temperature \cite{Tewordt-1963,Werthamer-1963}.

For these reasons, the LPDA approach may end up in producing physically reasonable profiles for $|\Delta(x)|$ and $\phi(x)$, provided that $\delta\phi$ is somewhat smaller than $\pi$. 
It may then be possible, in practice, to draw only an arc of the Josephson characteristics $J(\delta\phi)$ over a limited interval $0 \le \delta\phi \le \delta\phi_{\mathrm{max}}$ with $\delta\phi_{\mathrm{max}}$ somewhat smaller than $\pi$, rather than the complete Josephson characteristics over the whole interval $0 \le \delta\phi \le \pi$ as desirable.
This feature of the LPDA approach may also persist at finite temperature, although $\delta\phi_{\mathrm{max}}$ is expected to increase for increasing temperature. 

\begin{figure}[t]
\begin{center}
\includegraphics[width=8.7cm,angle=0]{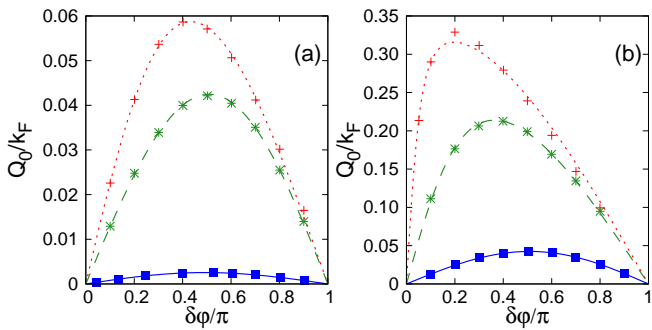}
\caption{(Color online) Josephson characteristics for a rectangular barrier of width $Lk_{F}= 4$ and various height $V_{0}/E_{F}$,
                                     where the values of $Q_{0}/k_{F}$ obtained in Ref.~\cite{SPS-2010} by solving the BdG equations at zero temperature for various couplings (symbols) are  
                                     compared with the fits obtained by Eq.~(\ref{universal-fitting-function}) (lines).
                                     In (a) $V_{0}/E_{F} = 0.4$, with $(k_{F}a_{F})^{-1} = -1$ (crosses and dotted line) where $A = 0.074$ and $B = 1.3$,
                                                                                    $(k_{F}a_{F})^{-1} = 0$ (stars and dashed line) where $A = 0.0412$ and $B = - 0.11$,
                                                                             and $(k_{F}a_{F})^{-1} = + 1.5$ (squares and full line) where $A = 0.00261$ and $B = 0.1$.
                                     In (b) $(k_{F}a_{F})^{-1} = 0$, with $V_{0}/E_{F} = 0.025$ (crosses and dotted line) where $A = 1.72$ and $B = 96$,
                                                                                    $V_{0}/E_{F} = 0.1$ (stars and dashed line) where $A = 0.367$ and $B = 4.9$,
                                                                             and $V_{0}/E_{F} = 0.4$ (squares and full line) where $A = 0.0412$ and $B = - 0.11$.}
\label{Figure-1}
\end{center}
\end{figure} 

Under these circumstances, we shall find it convenient to utilize a universal function of the form 
\begin{equation}
F(\delta\phi) = \frac{A \, \sin(\delta\phi)}{\sqrt{1+B \sin^{2}(\delta\phi/2)}} \, ,
\label{universal-fitting-function}
\end{equation}
where $A$ and $B$ are (dimensionless) numerical parameters that depend on the barrier as well as on coupling and temperature, in order to fit the curves $|\mathbf{Q}_{0}|/k_{F}$ vs $\delta\phi$ (or $J/J_{F}$ vs $\delta\phi$ where $J_{F} = k_{F} n/m$) that we shall obtain by the LPDA approach.
And this in spite of the fact that these curves may extend over a limited interval $0 \le \delta\phi \le \delta\phi_{\mathrm{max}}$ as discussed above.
The expression (\ref{universal-fitting-function}) satisfies the requirements $F(\delta\phi) = 0$ for $\delta\phi = 0$ and $\delta\phi = \pi$ \cite{Barone-1982}, and can be utilized only for $0 \le \delta\phi \le \pi$.
In addition, when $B=0$ it recovers the standard form $A \, \sin(\delta\phi)$ originally obtained for the Josephson effect \cite{Josephson-1962,Mahan-2000};
when $B \gg 1$ it reduces to $(2 A / \sqrt{B}) \cos(\delta\phi /2)$ for $0 < \delta\phi \le \pi$; and when $B \rightarrow - 1$ it reduces to $2 A \sin(\delta\phi /2)$ for $0 \le \delta\phi < \pi$.
For an arbitrary value of $B$ in the interval $-1 \le B \le + \infty$, the maximum of $F(\delta\phi)$ occurs at $\cos(\delta\phi) = ( 2 + B - 2 \sqrt{1 + B} ) / B$.
An expression of the type (\ref{universal-fitting-function}) was considered in the BCS limit for a tunnel junction, although only for negative values of $B$ \cite{Golubov-2004}.

In Sec.~\ref{sec:Josephson-characteristics} several examples will be reported, showing how the Josephson characteristics obtained by the LPDA approach recover the above limiting conditions, depending  
on the ranges of coupling, temperature, and barrier height and width.
To test the validity of the expression (\ref{universal-fitting-function}), we have utilized it to fit the Josephson characteristics that were obtained in Ref.~\cite{SPS-2010} by solving the BdG 
equations at zero temperature, for several barriers and various couplings throughout the BCS-BEC crossover.
The BdG numerical data from Ref.~\cite{SPS-2010} (symbols) and the corresponding fits (lines) are shown in Fig.~\ref{Figure-1}, where the values 
of the coefficients $A$ and $B$ are also reported in each case.
The good agreement shown in this figure over a wide range of values of $A$ and $B$ gives us confidence for utilizing the expression (\ref{universal-fitting-function}) to extrapolate 
the numerical data for the Josephson characteristics obtained by the LPDA approach, beyond $\delta\phi_{\mathrm{max}}$ and even up to $\delta\phi = \pi$.

\begin{figure}[t]
\begin{center}
\includegraphics[width=8.7cm,angle=0]{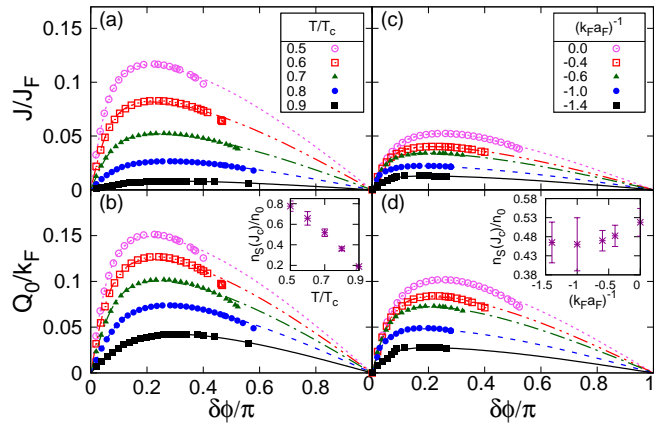}
\caption{(Color online): The Josephson characteristics for the current $J$ (in units of $J_{F}=k_{F} n/m$) 
              (panels (a) and (c)) are compared with those for the wave vector $Q_{0}=|\mathbf{Q}_{0}|$ (in units of $k_{F}$) (panels (b) and (d)), for a barrier of height $V_{0}/E_{F} = 0.05$ and width $k_{F}L = 3$.
              The left panels show the dependence of the Josephson characteristics on temperature for $(k_{F} a_{F})^{-1} = 0$, while the right panels show their dependence 
              on coupling for $T/T_{c} = 0.7$.
              The insets in the lower panels show the temperature and coupling dependence of $n_{\mathrm{s}}/n$ (see the text).}
\label{Figure-2}
\end{center}
\end{figure} 

\section{Josephson characteristics at finite temperature} 
\label{sec:Josephson-characteristics}

In this Section we provide a detailed description of the Josephson current-phase characteristics $J(\delta \phi)$ in terms of the LPDA approach described in
Sec.~\ref{sec:theoretical_approach}, over an extended region of the temperature-coupling phase diagram of the Fermi system, with barriers of different heights and widths.
To this end, whenever necessary we shall adopt the procedure discussed in Sec.~\ref{sec:theoretical_approach}-C, so as to draw the Josephson characteristics over the whole interval
$0 \le \delta\phi \le \pi$.
For simplicity, a rectangular barrier centered about $x=0$ of height $V_{0}$ and width $L$ will be utilized  in the following.
We shall consider either the dependence of the Josephson characteristic on temperature and coupling for given barrier, or their dependence on the barrier height and width for given temperature and coupling.

\vspace{0.05cm}
\begin{center}
{\bf A. Dependence of the Josephson characteristics on temperature and coupling}
\end{center}

Figure~\ref{Figure-2} shows how the Josephson characteristics $J(\delta \phi)$ evolve, as a function of temperature for given coupling [panel (a)] and as a function
of coupling for given temperature [panel (c)] with a fixed barrier.
In particular, from panel (a) the maximum value $J_{c}$ of $J$ is seen to decrease for increasing temperature, while the corresponding value of $\delta\phi$ approaches $\pi/2$ as 
$T \rightarrow T_{c}^{-}$ in such a way that $J(\delta \phi) \propto \sin(\delta\phi)$ (cf. the expression (\ref{universal-fitting-function}) with $B \rightarrow 0$).
In addition, panel (c) shows that $J_{c}$ decreases from unitarity to the BCS side of the crossover.
A systematic investigation of the Josephson critical current $J_{c}$ as a function of  temperature and coupling for given barrier is postponed to Sec.~\ref{sec:critical-current}.

\begin{figure}[t]
\begin{center}
\includegraphics[width=8.7cm,angle=0]{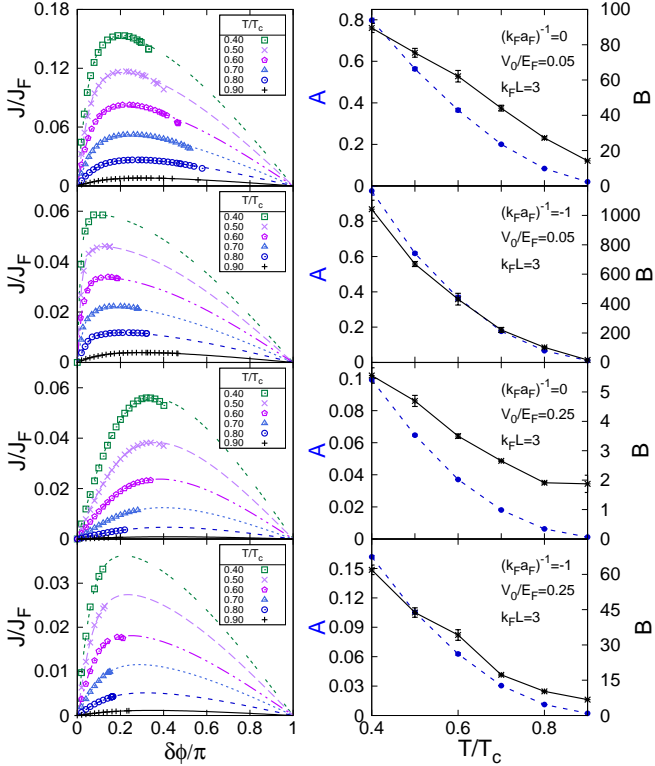}
\caption{(Color online): Left panels: Temperature dependence of the Josephson characteristics for the couplings $(k_{F} a_{F})^{-1} = 0$ and $(k_{F} a_{F})^{-1} = -1$, 
               with two rectangular barriers of heights $V_{0}/E_{F} = (0.05,0.25)$ and width $k_{F}L = 3$.
               Right panels: Corresponding temperature dependence of the coefficients $A$ (dots) and $B$ (stars) of the expression (\ref{universal-fitting-function}) used to fit the 
               corresponding Josephson characteristics in the left panels.}
\label{Figure-3}
\end{center}
\end{figure} 

For comparison, panels (b) and (d) of Fig.~\ref{Figure-2} show the corresponding behavior of the magnitude $Q_{0}$ of the wave vector $\mathbf{Q}_{0}$ 
that enters the expression (\ref{two-fluid-current}) for the current.
With the help of Eq.~(\ref{current-superfluid-density}), sufficiently away from the barrier this expression can be cast in the compact form 
$\frac{J}{J_{F}} = \frac{n_{\mathrm{s}}}{n} \, \frac{Q_{0}}{k_{F}}$ in terms of the superfluid density $n_{\mathrm{s}}$ which depends itself on $Q_{0}$. 
In particular, at $T = 0$ $n_{\mathrm{s}} = n$ irrespective of $Q_{0}$, such that $J/J_{F} = Q_{0}/k_{F}$ for a stable superfluid flow.
For increasing $T$, on the other hand, $n_{\mathrm{s}} < n$ such that $J/J_{F} < Q_{0}/k_{F}$.
This feature clearly emerges when comparing the curves of the bottom panels with those of the upper panels of Fig.~\ref{Figure-2}.
From this comparison, the ratio $n_{\mathrm{s}}/n$ can be extracted in each case for given $Q_{0}/k_{F}$.
The insets in the lower panels show the ratio $n_{\mathrm{s}}/n$ obtained for the values of $Q_{0}/k_{F}$ at which the Josephson characteristics have their maximum.
As expected, the ratio $n_{\mathrm{s}}/n$ obtained in this way is a decreasing function of temperature for given coupling and an increasing function of coupling for given temperature.

\begin{figure}[t]
\begin{center}
\includegraphics[width=8.7cm,angle=0]{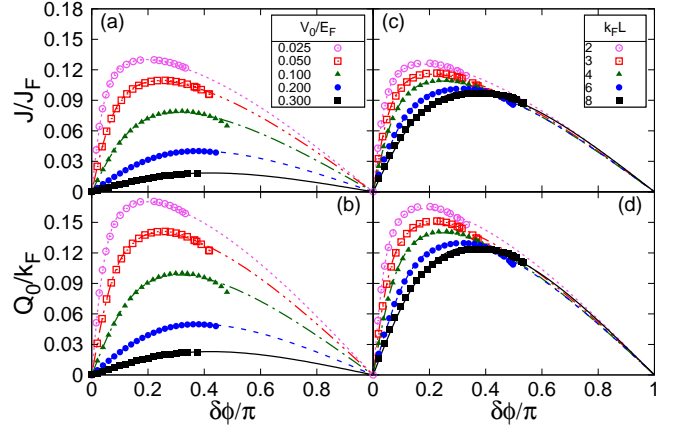}
\caption{(Color online): The Josephson characteristics for the current $J$ (in units of $J_{F}$) (panels (a) and (c)) are compared with those for the wave vector 
               $Q_{0}=|\mathbf{Q}_{0}|$ (in units of $k_{F}$) (panels (b) and (d)) for $T/T_{c} = 0.5$ and $(k_{F}a_{F})^{-1}= 0$, for barriers with several heights and widths.
               In particular, the left panels show the dependence of the Josephson characteristics on the height $V_{0}/E_{F}$ when $k_{F}L= 4$, while the right panels show their dependence 
               on $k_{F}L$ for $V_{0}/E_{F}=0.05$.}
\label{Figure-4}
\end{center}
\end{figure} 

It is interesting to compare the temperature dependence of the Josephson characteristics for the couplings $(k_{F} a_{F})^{-1} = 0$ and $(k_{F} a_{F})^{-1} = -1$, which are representative of the crossover 
region and the BCS regime, respectively.
This comparison is shown in Fig.~\ref{Figure-3} for two different barriers, together with the temperature dependence of the coefficients $A$ and $B$ that enter the fitting function 
(\ref{universal-fitting-function}) for the Josephson characteristics.
In all cases, note that for $\delta \phi \lesssim 0.1$ the Josephson characteristics in the BCS regime have a steeper dependence on $\delta \phi$ than those at unitarity.
This feature, in turn, is reflected in the values of the coefficient $B$, which are about ten times larger on the BCS side than at unitarity for both barriers.
On the other hand, the values of the coefficient $A$ remain comparable in the two cases.

\begin{figure}[t]
\begin{center}
\includegraphics[width=8.7cm,angle=0]{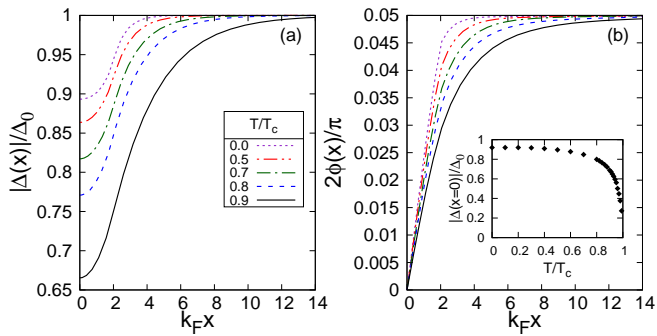}
\caption{(Color online): Spatial profiles of (a) $|\Delta(x)|$ (in units of the bulk value $\Delta_{0}$ at the given temperature in the presence of the current) 
               and (b) $2\phi(x)$, for $(k_{F} a_{F})^{-1} = 0$ and several temperatures in the interval $0 \le T/T_{c} \le 0.9$.
               The barrier has height $V_{0}/E_{F} = 0.05$ and width $k_{F}L= 4$, and the value $\delta \phi/\pi = 0.10$ is kept fixed for all profiles.
               Symbols are common to both panels.
               The inset shows the temperature dependence of $|\Delta(x=0)|/\Delta_{0}$ when $T$ approaches $T_{c}$.}
\label{Figure-5}
\end{center}
\end{figure} 

\vspace{0.05cm}
\begin{center}
{\bf B. Dependence of the Josephson characteristics on barrier height and width}
\end{center}

One can also examine the dependence of the Josephson characteristics on the barrier height and width, for given coupling and temperature.
An example is reported in Fig.~\ref{Figure-4}, where the left panels show the dependence on the height $V_{0}$ for given width $L$ and the right panels show the dependence on the width $L$ for given height $V_{0}$. 
In particular, from the left panels of Fig.~\ref{Figure-4} one sees that, for $T$ sufficiently smaller than $T_{c}$ and $V_{0} << E_{F}$, the Josephson characteristics tend to a $\cos (\delta \phi/2)$ dependence, 
while for increasing $V_{0}$ they acquire the standard $\sin (\delta \phi)$ dependence of the Josephson regime \cite{Mahan-2000} (which in Fig.~\ref{Figure-2} was seen to be anyway recovered when $T$ approaches $T_{c}$).
In addition, from the right panels of Fig.~\ref{Figure-4} one sees that, for given temperature and increasing $L$, the position of the maximum of the characteristic shifts to larger values
of $\delta \phi$, while the value of the maximum itself approaches a limiting value \cite{Zou-Dalfovo-2014}.
The physical implications of these results will be discussed at length in Sec.~\ref{sec:spatial-profiles-in}, even under more extreme circumstances that may give rise to a re-entrant behaviour of the Josephson characteristics \cite{Golubov-2004}.

\section{Spatial profiles of the magnitude and phase of the gap parameter} 
\label{sec:spatial-profiles-out}

In this Section, we consider the spatial profiles of the (complex) gap parameter $\Delta(x) = |\Delta(x)| e^{i 2 \phi(x)}$, and determine how $|\Delta(x)|$ and $2\phi(x)$ depend on temperature and coupling as well as on the height and width of the barrier.
From these profiles, we will extract the healing length $\xi_{\mathrm{out}}$ past which $|\Delta(x)|$ and $2\phi(x)$ approach their bulk values sufficiently far away from the barrier.
Results will be shown for $x \ge 0$ only, since $|\Delta(x)|$ is even and $\phi(x)$ is odd about $x=0$ within the chosen geometry.

\vspace{0.05cm}
\begin{center}
{\bf A. Dependence of the spatial profiles \\ on temperature and coupling}
\end{center}

The spatial profiles of the magnitude $|\Delta(x)|$ and phase $2\phi(x)$ of the gap parameter are reported in Fig.~\ref{Figure-5} at unitarity for several temperatures,
ranging from $T=0$ to $T=0.9 T_{c}$.
As expected, the spatial spreads of $|\Delta(x)|$ and $2\phi(x)$ increase for increasing temperature at given coupling.
Note also from the inset that the value of $|\Delta(x=0)|/\Delta_{0}$ at the center of the barrier rapidly decreases as $T \rightarrow T_{c}^{-}$, eventually approaching zero in the limit.
We have verified that these features hold for all couplings in the interval $-1.5 \lesssim (k_{F} a_{F})^{-1} \lesssim 0$ on the BCS side of unitarity, in which we have mostly concentrated our numerical efforts.

\begin{figure}[t]
\begin{center}
\includegraphics[width=8.7cm,angle=0]{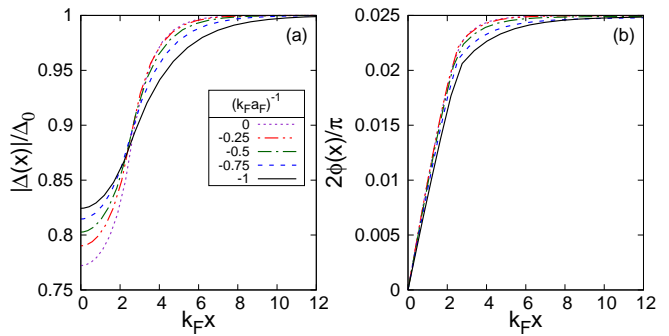}
\caption{(Color online): Spatial profiles of (a) $|\Delta(x)|$ (in units $\Delta_{0}$) and (b) $2\phi(x)$, 
               for $T/T_{c} = 0.5$ and several couplings on the BCS side of unitarity.
               The barrier has height $V_{0}/E_{F} = 0.1$ and width $k_{F}L= 5$, and the value $\delta \phi/\pi = 0.05$ is kept fixed for all profiles.
               Symbols are common to both panels.}
\label{Figure-6}
\end{center}
\end{figure} 

Figure~\ref{Figure-6} shows instead the spatial profiles of $|\Delta(x)|$ and $2\phi(x)$ at fixed value of $T/T_{c}$ for several couplings on the BCS side of unitarity.
For increasing coupling at given temperature, the depression $|\Delta(x=0)|/\Delta_{0}$ increases while the spatial spread of $|\Delta(x)|/\Delta_{0}$ and $2\phi(x)$ decreases, as also expected.
A similar effect of the same origin was reported in Refs.~\cite{Shanenko-2010,Shanenko-2012} in a related context, where (geometrical) quantum size effects played the role of the increasing inter-particle coupling discussed here.
Note also from Fig.~\ref{Figure-6} that all curves of $|\Delta(x)|/\Delta_{0}$ cross each other for $x$ slightly larger than $L/2$, where the curvature of the curves changes sign.
We have verified that these features remain true for all barrier heights and widths that we have considered, in the region of the temperature-coupling phase diagram where the LPDA approach 
can be applied with confidence (namely, $0.3 \lesssim T/T_{c}$ and $-1 \lesssim  (k_{F} a_{F})^{-1} \lesssim 0$), for the reasons discussed on physical grounds in the Introduction.

\vspace{0.05cm}
\begin{center}
{\bf B. Dependence of the the spatial profiles \\ on barrier height and width}
\end{center}

\begin{figure}[t]
\begin{center}
\includegraphics[width=8.7cm,angle=0]{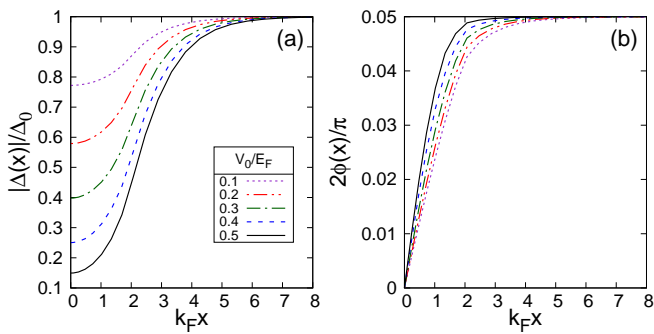}
\caption{(Color online): Spatial profiles of (a) $|\Delta(x)|$ (in units of $\Delta_{0}$) and (b) $2\phi(x)$, for a barrier 
               of width $k_{F}L= 4$ and several heights $V_{0}/E_{F} = (0.1,0.2,0.3,0.4,0.5)$. 
               Here, $(k_{F} a_{F})^{-1} = 0$ and $T/T_{c} = 0.5$, while the value $\delta \phi/ \pi = 0.1$ is kept fixed in all profiles.
               Symbols are common to both panels.}
\label{Figure-7}
\end{center}
\end{figure} 

A typical example of the dependence of $|\Delta(x)|$ and $2\phi(x)$ on the barrier height is shown in Fig.~\ref{Figure-7}, at unitarity and for given value of $T/T_{c}$.
Note, in particular, how the depression of $|\Delta(x)|$ at the center of the barrier becomes rapidly more pronounced for increasing $V_{0}$.

\begin{figure}[t]
\begin{center}
\includegraphics[width=8.7cm,angle=0]{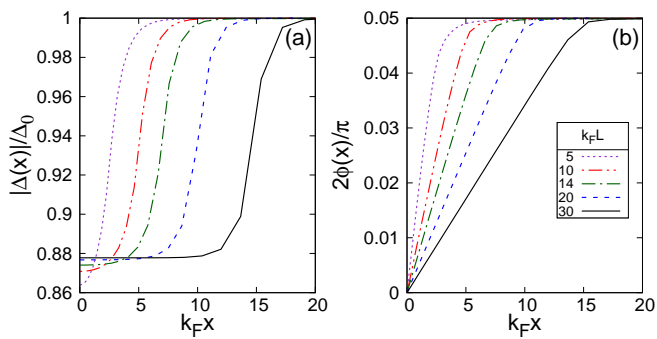}
\caption{(Color online): Spatial profiles of (a) $|\Delta(x)|$ (in units of $\Delta_{0}$) and (b) $\phi(x)$, for a barrier 
               of height $V_{0}/E_{F} = 0.05$ and several widths $k_{F}L= (5,10,14,20,30)$.
               Here, $(k_{F} a_{F})^{-1} = 0$ and $T/T_{c} = 0.5$, while the value $\delta \phi/ \pi = 0.1$ is kept fixed in all profiles.
               Symbols are common to both panels.}
\label{Figure-8}
\end{center}
\end{figure} 

The corresponding dependence of $|\Delta(x)|$ and $2\phi(x)$ on the barrier width is shown in Fig.~\ref{Figure-8}. 
Here, one notes that the value of $|\Delta(x=0)|$ is (almost) independent from the width $L$ while the spatial spread of $|\Delta(x)|$ and $2\phi(x)$ strongly depend on $L$.
One also notes that, for sufficiently large $L$, a region appears inside the barrier where $|\Delta(x)|$ becomes flat, which implies the emergence of a ``mini-gap'' inside this region.
We will consider this situation more extensively in Sec.~\ref{sec:spatial-profiles-in}.

\vspace{0.05cm}
\begin{center}
{\bf C. Extracting the healing length $\xi_{\mathrm{out}}$ from \\ the spatial profiles of the gap parameter}
\end{center}

It is worth characterizing in a systematic way the spread of the profiles $|\Delta(x)|$ and $2\phi(x)$ as shown in Figs.~\ref{Figure-5}-\ref{Figure-8}, as functions of all parameters at our disposal.

To this end, from a given spatial profile of $|\Delta(x)|$ we can extract the healing length $\xi_{\mathrm{out}}$ for restoring the bulk value of the gap parameter \emph{outside\/} the barrier, 
by considering an exponential fit of the type
\begin{equation}
|\Delta(x)|/\Delta_{0} = \left( 1 - C_{\mathrm{out}} \, e^{-(x - L/2)/\xi_{\mathrm{out}}} \right) \hspace{0.3cm} \mathrm{for} \hspace{0.3cm} x \gtrsim L/2
\label{exponential-fit-out}
\end{equation}
in terms of the parameters $C_{\mathrm{out}}$ and $\xi_{\mathrm{out}}$. 
The restriction to $x \gtrsim L/2$ reflects the fact that the curvature of $|\Delta(x)|/\Delta_{0}$ becomes negative for $x$ slightly larger than $L/2$, as noted after Fig.~\ref{Figure-6}.
An analogous procedure can be applied to the profiles of $\phi(x)$ or, better, of $d\phi(x)/dx$ to improve on the numerical precision.

\begin{figure}[t]
\begin{center}
\includegraphics[width=8.5cm,angle=0]{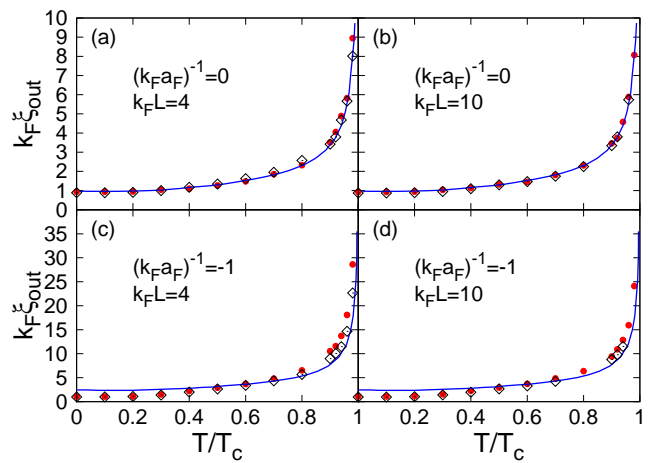}
\caption{(Color online): Healing length $\xi_{\mathrm{out}}$ (in units of $k_{F}^{-1}$) vs temperature (in units of $T_{c}$),
              obtained by fitting the numerical LPDA data with the expression (\ref{exponential-fit-out}). 
              Here, the barrier has height $V_{0}/E_{F} = 0.05$ (dots) and $V_{0}/E_{F} = 0.5$ (diamonds), and width $k_{F}L= 4$ (left panels) and $k_{F}L= 10$ (right panels), 
              while $\delta \phi = 0$ in all cases.
              Two coupling values $(k_{F}a_{F})^{-1} = 0$ (panels (a) and (b)) and $(k_{F}a_{F})^{-1}= -1$ (panels (c) and (d)) are considered. 
              A comparison is also shown with the temperature dependence of the healing length $\xi$ for a homogeneous system (full lines), whereby pairing fluctuations
              beyond mean field are included for the same couplings (cf. Fig.~10 of Ref.~\cite{Palestini-2014}).
              For this comparison, the results of the present calculation are rescaled by a factor $6/5$, which accounts for the different 
              definitions for $\xi_{\mathrm{out}}$ and $\xi$ in the two independent calculations.}
\label{Figure-9}
\end{center}
\end{figure}  

We begin by considering how the profiles of $|\Delta(x)|$, obtained for various couplings and barrier heights and widths by keeping $\delta \phi =0$, evolve as a function of temperature.
The corresponding values of the healing length $\xi_{\mathrm{out}}$ extracted from the exponential fit (\ref{exponential-fit-out}) are shown in Fig.~\ref{Figure-9} for a few characteristic cases.
It turns out that, when $\delta \phi = 0$, the values of $\xi_{\mathrm{out}}$ does not depend on the barrier height and width, but only on coupling and temperature (as it would be the case for a homogeneous system in the absence of the barrier).
In support of this finding, Fig.~\ref{Figure-9} shows also a comparison with the temperature dependence of the healing length $\xi$ for a homogeneous system, that was obtained in Ref.~\cite{Palestini-2014} by including pairing fluctuations beyond mean field.
A related agreement was reported in Ref.~\cite{SPS-2013}, between the results for the healing length of Ref.~\cite{Palestini-2014} and those obtained by solving the BdG equations for of an isolated vortex.
In Sec.~\ref{sec:Landau-criterion} we will again return to this connection, between the results obtained at the mean-field level in an inhomogeneous environment and those obtained with the further inclusion of pairing fluctuation in the homogeneous case.

\vspace{0.05cm}
\begin{center}
{\bf D. Spatial profiles and healing length of the gap parameter along the Josephson characteristics}
\end{center}

\begin{figure}[t]
\begin{center}
\includegraphics[width=8.7cm,angle=0]{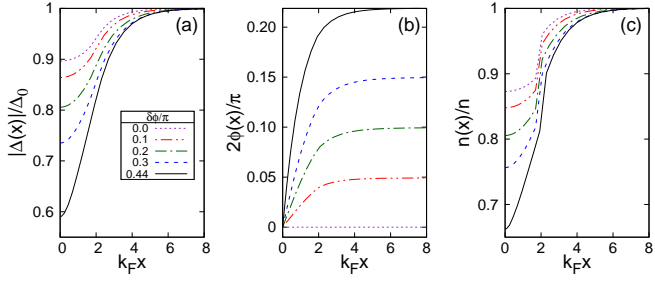}
\caption{(Color online): Spatial profiles of (a) $|\Delta(x)|$ (in units of $\Delta_{0}$), (b) $2\phi(x)$, and (c) the density $n(x)$ (in units of the bulk density $n$), 
               for a barrier of height $V_{0}/E_{F} = 0.05$ and width $k_{F}L= 4$. 
               Here, $(k_{F} a_{F})^{-1} = 0$ and $T/T_{c} = 0.5$, while several values of $\delta \phi/ \pi= (0.0,0.1,0.2,0.3,0.44)$ are considered.
               Symbols are common to all panels.}
\label{Figure-10}
\end{center}
\end{figure} 

Thus far, we have compared the spatial profiles of $|\Delta(x)|$ and $2\phi(x)$ by varying alternatively temperature, coupling, barrier height and width, while keeping fixed the
value of $\delta \phi$.
We pass now to compare these profiles for $\delta \phi$ proceeding along a Josephson characteristic.

Figure~\ref{Figure-10} shows typical profiles of $|\Delta(x)|$ and $2\phi(x)$ for an extended range of values of $\delta \phi/ \pi$.
In this case, the profiles of the density $n(x)$ are also shown.
Note that to larger values of $\delta \phi/ \pi$ there correspond larger depressions of $|\Delta(x=0)|/\Delta_{0}$ and $n(x=0)/n$, as expected.

\begin{figure}[t]
\begin{center}
\includegraphics[width=8.7cm,angle=0]{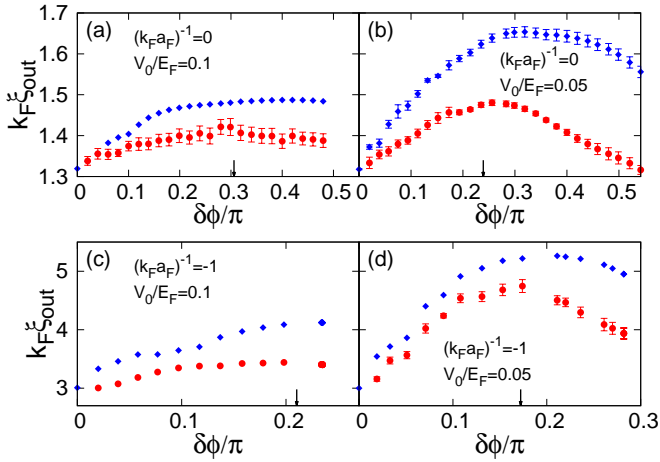}
\caption{(Color online): Healing length $\xi_{\mathrm{out}}$ (in units of $k_{F}^{-1}$) vs $\delta\phi$, at temperature $T/T_{c} = 0.7$ and coupling
              $(k_{F}a_{F})^{-1} = 0$ (panels (a) and (b)) and $(k_{F}a_{F})^{-1} = -1$ (panels (c) and (d)).
              Here, the barrier has width $Lk_{F}= 4$ and height $V_{0}/E_{F} = 0.1$ (left panels) and $V_{0}/E_{F} = 0.05$ (right panels). 
              Both the values of $\xi_{\mathrm{out}}$ extracted from the spatial profiles of $|\Delta(x)|$ (diamonds) and of $d \phi(x)/ dx$ (dots) are shown, 
              where the error bars arise from the fitting procedure.
              In each panel, the arrow points at the value of $\delta\phi / \pi$ where the corresponding Josephson characteristic has its maximum.}
\label{Figure-11}
\end{center}
\end{figure} 

From the profiles of $|\Delta(x)|$ and $2\phi(x)$ like those in Fig.~\ref{Figure-10}, we can extract the values of the healing length $\xi_{\mathrm{out}}$ along 
a Josephson characteristic, for given barrier, coupling, and temperature.
Representative results of this analysis are reported in Fig.~\ref{Figure-11}, where we show not only
the values of $\xi_{\mathrm{out}}$ extracted from the spatial profiles of $|\Delta(x)|$, but for comparison also those extracted from the spatial profiles of $d\phi(x)/dx$.
Note that, in contrast to Fig.~\ref{Figure-9} where $\xi_{\mathrm{out}}$ did not depend on the barrier height and width for $\delta \phi = 0$, 
in Fig.~\ref{Figure-11} the occurrence of this dependence appears as soon as $\delta \phi \ne 0$.

For all cases reported in Fig.~\ref{Figure-11}, note also that the numerical values of $\xi_{\mathrm{out}}$ extracted from the spatial profiles of $|\Delta(x)|$ somewhat differ from those extracted 
from the spatial profiles of $d\phi(x)/dx$ (which require $\delta \phi \ne 0$), although they both share the same overall shape as function of $\delta \phi$ with the maximum occurring at the same value of $\delta \phi$.
Note further that the value corresponding to the maximum is slightly larger than the value of $\delta \phi$ at which the corresponding Josephson characteristic has its maximum (indicated by an arrow in each panel).

\begin{figure}[t]
\begin{center}
\includegraphics[width=8.0cm,angle=0]{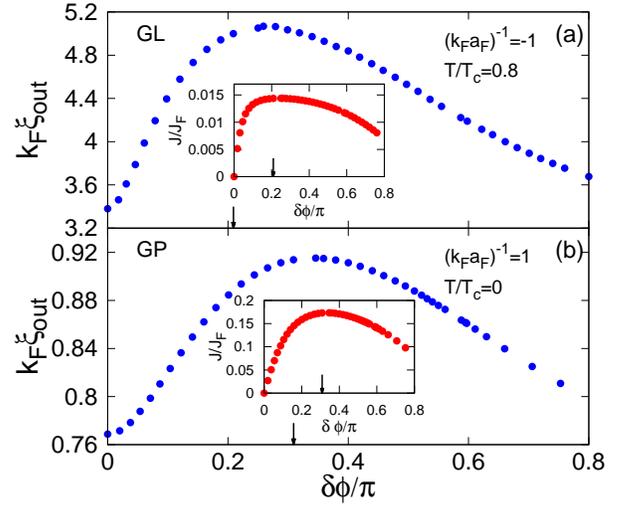}
\caption{(Color online): Healing length $\xi_{\mathrm{out}}$ (in units of $k_{F}^{-1}$) vs $\delta\phi$, extracted from the spatial profiles of $|\Delta(x)|$ 
              for a barrier of width $k_{F}L= 4$ and height $V_{0}/E_{F} = 0.05$.
              Here, $(k_{F}a_{F})^{-1} = -1$ and $T/T_{c} = 0.8$ (panel (a)) and $(k_{F}a_{F})^{-1} = +1$ and $T = 0$ (panel (b)), which correspond to the conditions when 
              the GL and GP equations can respectively be applied.
              The arrows point at the values of $\delta\phi / \pi$ where the corresponding Josephson characteristics (shown also for comparison in the insets) have the maximum.}
\label{Figure-12}
\end{center}
\end{figure} 

In this context, two opposite cases deserve separate consideration, namely, for weak coupling and $T \simeq T_{c}$ and for strong coupling and $T \simeq 0$, when the results obtained by the LPDA approach reduce, respectively, to those obtained by the GL and GP equations.
The results for $\xi_{\mathrm{out}}$ obtained in these cases from the spatial profiles of $|\Delta(x)|$ with a typical barrier are shown in Fig.~\ref{Figure-12} over a wide range of $\delta \phi/ \pi$, 
where also the corresponding Josephson characteristics are reported for comparison.
Even in these cases, note that the maximum of the curve $\xi_{\mathrm{out}}(\delta\phi)$ occurs at a slightly larger value of $\delta\phi$ with respect to the corresponding Josephson characteristics $J(\delta\phi)$.

Of the GL and GP cases which can both be obtained from the LPDA equation, in the following (see Sec.~\ref{sec:GL-regime}) we are going to give special attention to the GL case, by letting $T$ to approach $T_{c}$ from below
for couplings on the BCS side of unitarity.
And this not only because the LPDA equation (alike the BdG equations from which it originates), by not explicitly including the beyond-mean-field fluctuations required to span the entire temperature-coupling phase diagram of the
BCS-BEC crossover \cite{Physics-Reports-2018}, is expected to be appropriate at finite temperature especially on the BCS side of unitarity.
But also because of the significance that the GL equation has assumed for applications of superconductivity phenomena \cite{Tinkham-1980}.

\begin{figure}[t]
\begin{center}
\includegraphics[width=8.8cm,angle=0]{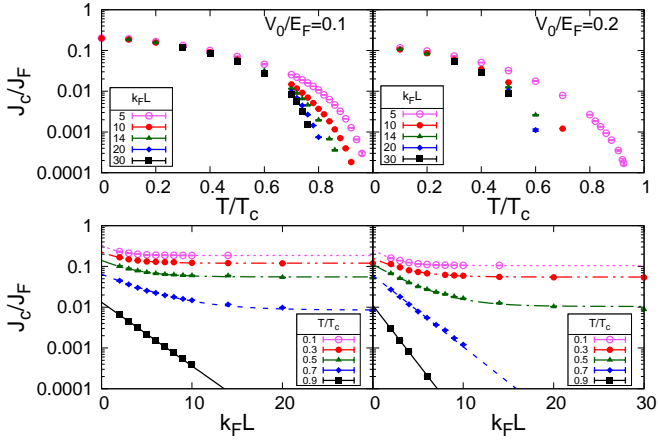}
\caption{(Color online): Upper panels: Temperature dependence of the critical current $J_{c}$ (in units of $J_{F}$) at unitarity for two barrier heights  and several barrier widths.
                                                                                            Lower panels: Width dependence of $J_{c}$ for several temperatures. The curves through the numerical data (symbols) correspond to fits
                                                                                            obtained by the expression (\ref{exponential-fit-in-Jc}).}
\label{Figure-13}
\end{center}
\end{figure} 

\section{Spatial profiles in the interior of the barrier and the emergence of the proximity effect} 
\label{sec:spatial-profiles-in}

In this Section, we consider the complementary healing length $\xi_{\mathrm{in}}$ that can be identified in the \emph{interior} of the barrier.
To extract it in a meaningful way, the barrier width $L$ of the internal (s) region of the SsS junction that we are considering will have to be sufficiently widened,  
so that an exponential fit analogous to Eq.~(\ref{exponential-fit-out}) becomes feasible.
In this way, we will be able to relate the shape of the gap profiles in the interior of the barrier to the appearance of a ``Josephson-induced proximity effect''.

\vspace{0.05cm}
\begin{center}
{\bf A. Behaviour of the critical current \\ for increasing barrier width}
\end{center}

We begin our analysis by following closely the experimental procedure of Ref.~\cite{Deutsher-1991} for an SNS junction, to extract the coherence length $\xi_{N}$ in the normal (N) region.
We thus look at the behaviour of the critical current $J_{c}$ when the width $L$ of the internal (s) region is progressively enlarged.
This procedure is then repeated for several temperatures in the range $0 \le T \le T_{c}$.

Figure~\ref{Figure-13} shows the results of this kind of analysis, made here at unitarity for two different barrier heights.
In particular, the upper panels show the temperature dependence of $J_{c}$ for several barrier widths, while the lower panels show the width dependence of $J_{c}$ for several temperatures 
(in analogy with Figs.~2 and 3 of Ref.\cite{Deutsher-1991}, respectively).
Note from the upper panels of Fig.~\ref{Figure-13} that the decrease of $J_{c}$ for increasing temperature can be followed over three orders of magnitude, and that this drop becomes 
more abrupt for increasing width $L$.
In addition, from the lower panels of Fig.~\ref{Figure-13} a definite behaviour of $J_{c}$ vs $L$ can be identified through a fit of the type
\begin{equation}
J_{c}(L) = J_{c}^{0} \, e^{-L/\xi_{\mathrm{in}}^{J_{c}}} + J_{c}^{\mathrm{mini}}
\label{exponential-fit-in-Jc}
\end{equation}
\noindent
performed in terms of three parameters $(J_{c}^{0},\xi_{\mathrm{in}}^{J_{c}},J_{c}^{\mathrm{mini}})$ over the extended range $0.1 \lesssim k_{F} L \lesssim 30$.
The results of these fits are represented by the curves in the lower panels of Fig.~\ref{Figure-13},
where to the curves that become flat for sufficiently large $L$ there corresponds a finite value of
the parameter $J_{c}^{\mathrm{mini}}$ (which, in turn, is associated with the existence of \emph{minigap\/} $\Delta_{\mathrm{mini}}$ in the internal (s) region of the SsS junction - see below).

\begin{figure}[t]  
\begin{center}
\includegraphics[width=8.6cm,angle=0]{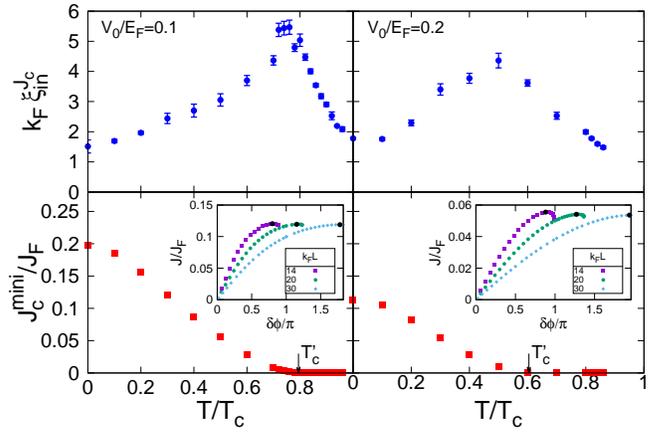}
\caption{(Color online): Upper panels: Temperature dependence of the healing length $\xi_{\mathrm{in}}^{J_{c}}$ (in units of $k_{F}^{-1}$) obtained at unitarity from the fit 
                                                                                         (\ref{exponential-fit-in-Jc}) for two barrier heights.
                                                                                         Lower panels: Corresponding temperature dependence of the parameter $J_{c}^{\mathrm{mini}}$ of the fit (\ref{exponential-fit-in-Jc}).
                                                                                         The vertical arrows mark the temperature $T_{c}^{'}$ where $J_{c}^{\mathrm{mini}}$ vanishes.
                                                                                         The insets show the Josephson characteristics for $T/T_{c}=0.3$ and three values of $k_{F}L$.}
\label{Figure-14}
\end{center}
\end{figure} 

The information extracted from the fits (\ref{exponential-fit-in-Jc}) is summarized in Fig.~\ref{Figure-14}, where the temperature dependence of the two parameters $\xi_{\mathrm{in}}^{J_{c}}$ and $J_{c}^{\mathrm{mini}}$  
is reported for two different barrier heights.
[We have also verified that, within numerical uncertainty, the value of $J_{c}(L \rightarrow 0) = J_{c}^{0} + J_{c}^{\mathrm{mini}}$ obtained from the expression (\ref{exponential-fit-in-Jc}) coincides with the value of
$J_{c}(0)$ that will be obtained in Sec.~\ref{sec:Landau-criterion} in the context of the Landau criterion for superfluidity.]
For completeness, the insets of Fig.~\ref{Figure-14} show the Josephson characteristics for three different barrier widths which, by sharing the same value of $J_{c}$ (black dots), signal 
the presence of a non-vanishing minigap.
Note how, for increasing $L$, the shape of the Josephson characteristics evolve toward a ``re-entrant'' (sometimes called ``multivalued'' \cite{Golubov-2004}) behaviour, whereby the value of $J_{c}$ 
occurs for $\delta \phi / \pi > 1$.
In this case, however, the fitting function (\ref{universal-fitting-function}) for the Josephson characteristics cannot be utilized.

What one clearly evidences from Fig.~\ref{Figure-14} are the strong enhancement of the healing length $\xi_{\mathrm{in}}^{J_{c}}$ and the corresponding vanishing of $J_{c}^{\mathrm{mini}}$ at a temperature  $T_{c}^{'}$, which is smaller than the critical temperature $T_{c}$ of the SsS junction and whose value depends on the barrier height.
Both these features are strongly reminiscent of the \emph{proximity effect} \cite{Deutscher-1969} as manifested, in particular, in an SNS junction \cite{Deutsher-1991}, whereby the leakage of Cooper pairs from the external (S) regions into the internal (N) region induces a superconducting behaviour in the N region and permits the flow of a stationary Josephson supercurrent.
Specifically, the behaviour shown in Fig.~\ref{Figure-14} for an SsS junction with large enough $L$ could be referred to as a \emph{``Josephson-induced proximity effect"}, which is in some sense complementary to what is already known in the literature as ``proximity-induced Josephson effect" \cite{Han-1985,TA-1988}.

\vspace{0.05cm}
\begin{center}
{\bf B. Emergence of the proximity effect and the effective coupling in the interior of the barrier}
\end{center}

We are in a position to characterize in more detail the features of the proximity effect that we have just discovered, by analysing the spatial profiles of the magnitude $|\Delta(x)|$ of the gap parameter in the internal (s) region inside the barrier.
For $L$ large enough, $|\Delta(x)|$ may develop a uniform \emph{minigap\/} $\Delta_{\mathrm{mini}}$ sufficiently deep in the internal region, such that for a given temperature below $T_{c}$ we have to consider an exponential 
fit of the type
\begin{equation}
|\Delta(x)| = \Delta_{\mathrm{in}} \, e^{-|x + L/2|/\xi_{\mathrm{in}}^{\Delta}} + \Delta_{\mathrm{mini}}  \hspace{0.4cm} \mathrm{for} \hspace{0.1cm} - L/2 \lesssim x \lesssim 0
\label{exponential-fit-in}
\end{equation}
in terms of two parameters $\Delta_{\mathrm{in}}$ and $\xi_{\mathrm{in}}^{\Delta}$. 
We recall that profiles of $|\Delta(x)|$ across a single SN interface were extensively obtained in Ref.~\cite{Piselli-2018} in terms of the non-local version of the LPDA equation with no superfluid flow.
Here, we obtain related profiles in terms of the LPDA equation, both in the absence and in the presence of a superfluid flow.
In this context, particular interest acquires the temperature dependence of the minigap $\Delta_{\mathrm{mini}}$ from Eq.~(\ref{exponential-fit-in}), since this quantity can be subject to experimental probe in terms of 
the local density of states \cite{Esteve-2008}.
Note also that $\xi_{\mathrm{in}}^{\Delta}$ can be extracted from Eq.~(\ref{exponential-fit-in}) in a meaningful way, provided that the results of the fit do not depend on the value of the barrier width $L$.
In practice, in the presence of a superfluid flow this can be obtained whenever $\Delta_{\mathrm{mini}}$ remains non-vanishing within numerical accuracy.
Otherwise the current flow through the SsS junction cannot be sustained numerically in a self-consistent way.

\begin{figure}[t]  
\begin{center}
\includegraphics[width=8.8cm,angle=0]{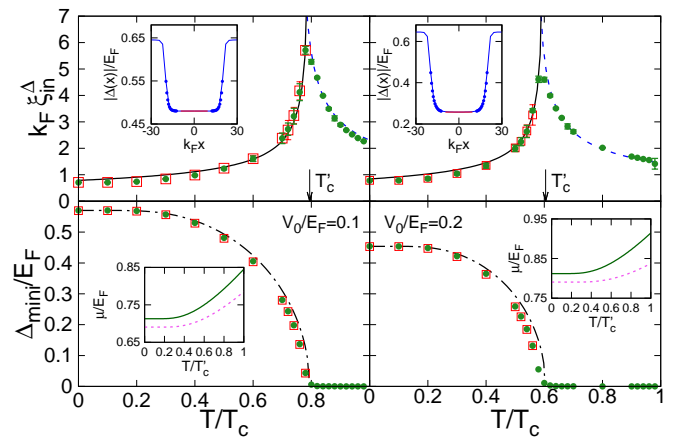}
\caption{(Color online): Upper panels: Temperature dependence of the healing length $\xi_{\mathrm{in}}^{\Delta}$ (in units of $k_{F}^{-1}$) obtained at unitarity from the fit 
                                                                                          (\ref{exponential-fit-in}) for two barrier heights.
                                                                                          The vertical arrows mark the temperature $T_{c}^{'}$ at which $\Delta_{\mathrm{mini}}$ vanishes.
                                                                                          The insets show the spatial profiles of $|\Delta(x)|$ in the presence of a current (lines) with the associated fits (dots) for $T/T_{c}=0.5$ and $k_{F}L=40$, 
                                                                                          from which the plateau corresponding to $\Delta_{\mathrm{mini}}$ can be identified.
                                                                                          Lower panels: Temperature dependence of the minigap $\Delta_{\mathrm{mini}}$ (with the same symbols of the upper panels).
                                                                                          The insets show the temperature dependence of the chemical potential at unitarity shifted upward by $V_{0}/E_{F}$ (broken lines) 
                                                                                          and of the chemical potential for the effective couplings listed in Table~\ref{Table-I} (full lines).
                                                                                          In all panels, the numerical data (symbols) are obtained for a given value of $\delta\phi$ in the Josephson characteristics, corresponding to 
                                                                                          zero current ($\delta\phi=0$ - filled dots) and a finite current ($\delta\phi/\pi=0.2$ - empty squares).
                                                                                          The meaning of the broken lines passing through these symbols is explained in the text.}
\label{Figure-15}
\end{center}
\end{figure} 

Figure~\ref{Figure-15} shows the temperature dependence of the healing length $\xi_{\mathrm{in}}^{\Delta}$ (upper panels) and of the minigap $\Delta_{\mathrm{mini}}$ (lower panels) 
for two values of the barrier height, as obtained from the fit (\ref{exponential-fit-in}) when applied to the numerical data obtained by the LPDA approach (symbols).
Both results, obtained with zero current ($\delta\phi=0$) and a finite value of the current (corresponding to $\delta\phi/\pi=0.2$), are shown in the figure.
As an internal check on the numerical calculations, we note that the values of $\xi_{\mathrm{in}}^{\Delta}$ from the upper panels of Fig.~\ref{Figure-15} and $\xi_{\mathrm{in}}^{J_{c}}$ 
from the upper panels of Fig.~\ref{Figure-14} show a similar behaviour, although they agree better with each other above $T_{c}^{'}$ where the fits for $\xi_{\mathrm{in}}^{J_{c}}$ become more reliable.

\begin{table}[h]
\begin{tabular}{c|c|c}
$V_{0}/E_{F}$ & $T_{c}^{\mathrm{'}}/T_{c}$ & $(k_{F}a_{F})^{-1}_\mathrm{{eff}}$ \\
\hline
0.1 & 0.794 & -0.189 \\
\hline
0.2 & 0.604 & -0.397 \\
\end{tabular}
\caption{Values of the temperature $T_{c}^{'}$ (in units of the bulk critical temperature $T_{c}$) at which $\Delta_{\mathrm{mini}}$ vanishes and of the related effective coupling $(k_{F}a_{F})^{-1}_\mathrm{{eff}}$,
               for the two barrier heights considered in Fig.~\ref{Figure-15}.}
\label{Table-I}
\end{table}

In addition, in the upper panels of Fig.~\ref{Figure-15} the lines passing through the symbols represent simple fits of the type $k_{F} \xi_{\mathrm{in}}^{\Delta} = C_{\mathrm{in}} \sqrt{1 - T/T_{c}^{'}}$ for $T < T_{c}^{'}$ and $k_{F} \xi_{\mathrm{in}}^{\Delta} = C_{\mathrm{in}}^{'} \sqrt{T/T_{c}^{'} -1}$ for $T > T_{c}^{'}$, where $T_{c}^{'}$ is the temperature at which $\Delta_{\mathrm{mini}}$ is seen to vanish from the numerical data reported 
in the lower panels of Fig.~\ref{Figure-15}.
From the value of $T_{c}^{'}$ obtained in this way, we then determine the values of the effective coupling $(k_{F}a_{F})^{-1}_\mathrm{{eff}}$ reported in Table~\ref{Table-I}, for which $T_{c}^{'}$ equals the critical temperature of a homogeneous Fermi superfluid along the BCS-BEC crossover.
The broken lines in the lower panels of Fig.~\ref{Figure-15} then correspond to the temperature dependence of the gap parameter $\Delta_{\mathrm{homo}}$ of this homogeneous Fermi superfluid with coupling $(k_{F}a_{F})^{-1}_\mathrm{{eff}}$ and no current flow.
The good agreement, that occurs without any fitting parameter between the temperature dependence of $\Delta_{\mathrm{homo}}$ and $\Delta_{\mathrm{mini}}$, gives us confidence about our interpretation 
of the results reported in Figs.~\ref{Figure-13} and \ref{Figure-14} in terms of a ``Josephson-induced proximity effect'', which acts to convert the internal (s) region of an SsS junction into the N region of an SNS junction with an appropriate effective coupling smaller than the coupling in the S region.
From a physical point of view, it is especially remarkable that the presence of a local one-body potential (the barrier) can locally induce a change in the two-body interaction.

Nevertheless, the limited range available for the width $L$ of the internal (s) region, that can be explored before the value of the critical current $J_{c}$ becomes too small to be numerically detectable, prevents the internal region to be fully identified with an independent piece of material with its own bulk thermodynamic properties.
Thermodynamic equilibrium between the external (S) regions with chemical potential $\mu$ and the internal (s) region inside the barrier of height $V_{0}$ with the \emph{effective} chemical potential 
$\mu_{\mathrm{eff}}$ associated with the coupling $(k_{F}a_{F})^{-1}_\mathrm{{eff}}$ would, in fact, require the identity $\mu_{\mathrm{eff}} - V_{0} = \mu$ to be satisfied.
As shown in the insets in the lower panels of Fig.~\ref{Figure-15}, however, this identity is only approximately satisfied by our numerical calculations, although the upward shift of $\mu$ by the amount $V_{0}$ makes the result quite close to $\mu_{\mathrm{eff}}$.
                                                                                
\section{Boundary of the Ginzburg-Landau \\ regime from the LPDA approach} 
\label{sec:GL-regime}

Strictly speaking, the GL approach is expected to be valid in the \emph{extreme\/} BCS limit $(k_{F}a_{F})^{-1} \rightarrow - \infty$ and for $T \rightarrow T_{c}^{-}$
\cite{Gorkov-1961}. 
It would then be of practical interest to somewhat relax these extreme conditions and identify an extended region in the temperature-coupling phase diagram below $T_{c}$
on the BCS side of unitarity, where the GL approach could be applied with confidence.
In this Section, we address this question by determining numerically the deviations of the GL approach from the LPDA approach, by exploiting the fact that, in principle, 
the latter is able to cover the whole superfluid sector of the temperature-coupling phase diagram.

\vspace{0.05cm}
\begin{center}
{\bf A. Analytic results \\ in the extreme BCS regime}
\end{center}

The GL equation, which we are going to confront with, has the form \cite{Physics-Reports-2018}:
\begin{eqnarray}
& & \frac{(i\mathbf{\nabla} + 2 \mathbf{A}(\mathbf{r}))^{2}}{4 \, m} \, \Delta(\mathbf{r}) + \frac{3}{4 E_F} \, |\Delta(\mathbf{r})|^2 \Delta(\mathbf{r}) 
\label{GL-equation} \\
& + & \frac{6 \pi^2 (k_{B}T_{c})^{2} }{7 \zeta(3) E_F} \!\! \left[ \! \left(1- \frac{\pi}{4k_Fa_F}\right) \!\! \frac{V(\mathbf{r})}{E_F} 
- \! \left(1 - \frac{T}{T_c} \right) \! \right] \! \Delta(\mathbf{r}) = 0
\nonumber
\end{eqnarray}
where $\zeta(3) \simeq 1.202$ is the Riemann $\zeta$ function of argument $3$ and the vector potential $\mathbf{A}(\mathbf{r})$ is identified with $- \mathbf{Q}_{0}$ in the context 
of the Josephson effect (cf. Sec.~\ref{sec:theoretical_approach}-A).
Note that the term in Eq.~(\ref{GL-equation}) containing the external potential $V(\mathbf{r})$ was not considered in the original Gor'kov derivation of the GL equation from the BdG equations \cite{Gorkov-1961} and was only later included in Ref.~\cite{Petrov-1998}.

The GL equation in the form (\ref{GL-equation}) (including the external potential $V(\mathbf{r})$) can be recovered from the LPDA equation (\ref{LPDA-equation}), by taking the appropriate limits 
$(k_{F}a_{F})^{-1} \ll -1$ and $T \rightarrow T_{c}^{-}$ in the coefficients $\mathcal{I}_{0}(\mathbf{r})$ and $\mathcal{I}_{1}(\mathbf{r})$ given by the expressions (\ref{I_0}) and (\ref{I_1}), respectively.
In these limits, for small enough $\mathbf{A}(\mathbf{r})$ one obtains:
\begin{eqnarray}
\mathcal{I}_{0}(\mathbf{r}) \cong & - & \frac{m}{4 \pi a_{F}} + N_{0} \! \left[ \!\! \left(1 - \frac{T}{T_{c}}\right) - \left(1 - \frac{\pi}{4 k_{F} a_{F}}\right) \!\! \frac{V(\mathbf{r})}{E_{F}} \! \right]
\nonumber \\
& - & N_{0} \frac{7 \, \zeta(3)}{8 \pi^{2} (k_{B}T_{c})^{2}} |\Delta(\mathbf{r})|^{2}  - \frac{\mathbf{A}(\mathbf{r})^{2}}{m} \, \mathcal{I}_{1}(\mathbf{r})
\label{I_0-GL-limit}
\end{eqnarray}
and 
\begin{equation} 
\mathcal{I}_{1}(\mathbf{r}) \cong N_{0} \, \frac{ 7 \zeta(3) E_{F}}{6 \pi^{2} (k_{B}T_{c})^{2}} 
\label{I_1-GL-limit}
\end{equation}
where $N_{0} = m k_{F}/(2 \pi^{2})$ is the density of states at the Fermi level per spin component.
The corresponding expression for the superfluid current is given by Eq.~(\ref{BCS-superfluid-current}), which is now imposed as a local constraint for the solution of the GL equation (\ref{GL-equation}) 
as we did in Eq.~(\ref{local-condition}).

\vspace{0.05cm}
\begin{center}
{\bf B. Numerical results \\ in an extended BCS regime}
\end{center}

\begin{figure}[t]  
\begin{center}
\includegraphics[width=8.6cm,angle=0]{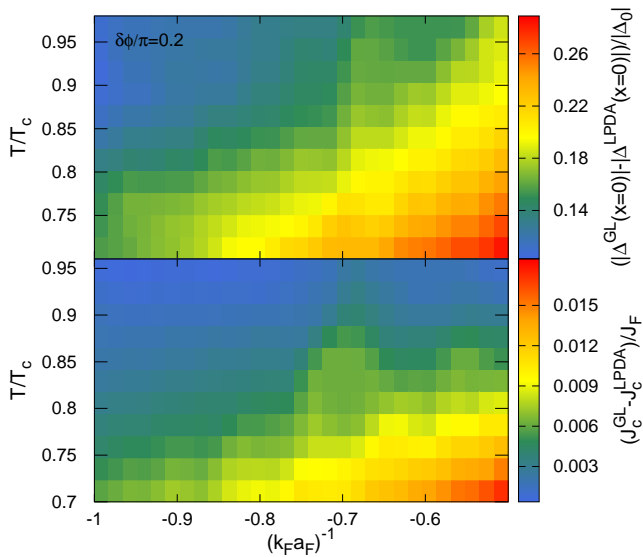}
\caption{(Color online): Color maps in the temperature-coupling phase diagram, showing the difference of the magnitude of the gap parameter $|\Delta(x=0)|$ at the barrier center 
                                     in units of the bulk value $\Delta_{0}$ (upper panel) and of the critical current $J_{c}$ in units of $J_{F}$ (lower panel), which are obtained by solving alternatively the GL equation (\ref{GL-equation}) 
                                     and the LPDA equation (\ref{LPDA-equation}).
                                     All numerical values are obtained with a barrier of height $V_{0}/E_{F }=0.1$ and width $k_{F}L =4$. 
                                     In addition, in the upper panel a fixed value $\delta \phi/\pi = 0.2$ is used.}
\label{Figure-16}
\end{center}
\end{figure} 

As a  preliminary check on how the LPDA approach smoothly evolves into the GL approach in the appropriate sector of the temperature-coupling phase diagram, we have monitored how the numerical values of the coefficients $\mathcal{I}_{0}(\mathbf{r})$ and $\mathcal{I}_{1}(\mathbf{r})$ obtained from the general expressions (\ref{I_0}) and (\ref{I_1}) progressively evolve towards those obtained from the limiting expressions (\ref{I_0-GL-limit}) and (\ref{I_1-GL-limit}).

It appears, however, more interesting from a physical point of view to determine how the profiles of the magnitude of the gap parameter $|\Delta(x)|$ as well as the values of the critical current $J_{c}$ evolve from the LPDA 
to the GL approach.
To this end, the upper panel of Fig.~\ref{Figure-16} shows the color map of the difference $(|\Delta^{\mathrm{GL}}(x=0)| - |\Delta^{\mathrm{LPDA}}(x=0)|)$ obtained at the center of the barrier 
(in units of the bulk value $\Delta_{0}$), while the lower panel of Fig.~\ref{Figure-16} shows the color map of the difference $(J_{c}^{\mathrm{GL}} - J_{c}^{\mathrm{LPDA}})$ (in units of $J_{F}$).
Both quantities are here reported in the relevant sector of the temperature-coupling phase diagram, where the LPDA results are explicitly seen to smoothly merge into the GL results.
For both quantities, this merging appears to be essentially complete when $(k_{F} a_{F})^{-1} \lesssim -0.7$ and $T/T_{c} \gtrsim 0.85$. 
As an outcome of our analysis, it then appears that the conditions for the applicability of the GL approach can indeed be somewhat relaxed from the extreme BCS regime into an extended BCS regime, 
as far as both coupling and temperature ranges are concerned.

\section{Critical current} 
\label{sec:critical-current}

The behaviour of the critical current $J_{c}$ associated with the Josephson characteristics was considered in Sec.~\ref{sec:spatial-profiles-in} at unitarity for large barrier widths, to show specifically the emergence of proximity effects.
In this Section, we consider the behaviour of $J_{c}$ in a more systematic way, for different barrier heights and widths, as well as couplings and temperatures.
Specific efforts will be made to determine the temperature dependence of $J_{c}$ upon approaching the critical temperature under various circumstances.

\vspace{0.05cm}
\begin{center}
{\bf A. Coupling dependence of the critical current}
\end{center}

\begin{figure}[t]
\begin{center}
\includegraphics[width=8.7cm,angle=0]{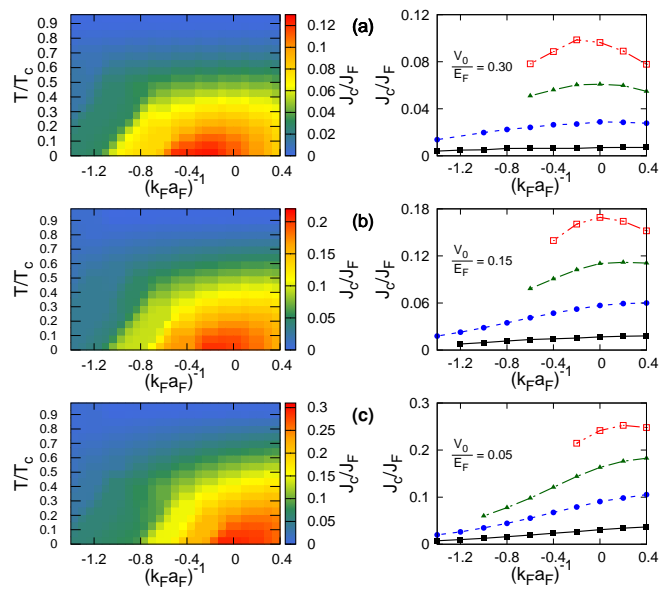}
\caption{(Color online): Critical current $J_{c}$ (in units of $J_{F}$) for a barrier of width $k_{F}L= 2$ and various heights: 
              (a) $V_{0}/E_{F} = 0.30$, (b) $V_{0}/E_{F} = 0.15$, and (c) $V_{0}/E_{F} = 0.05$.
              For each barrier, the left panels show the colour maps of $J_{c} / J_{F}$ in the temperature-coupling phase diagram, while the right panels show the curves $J_{c} / J_{F}$ vs $(k_{F}a_{F})^{-1}$
              for various temperatures: $T/T_{c} = 0.2$ (empty squares), $T/T_{c} = 0.4$ (filled triangles), $T/T_{c} = 0.6$ (filled circles), and $T/T_{c} = 0.8$ (filled squares).
              Note the different vertical scale in each panel.}
\label{Figure-17}
\end{center}
\end{figure} 

Figure~\ref{Figure-17} shows the colour maps of the critical current $J_{c}$ in the temperature-coupling phase diagram for a barrier of fixed width and decreasing heights
from top to bottom (left panels), together with the corresponding coupling dependence of $J_{c}$ for a choice of temperatures (right panels).
Note from this figure that the maximum value of $J_{c}$ shifts to weaker couplings toward the BCS side for increasing barrier height, but shifts back to stronger couplings toward the BEC side
for increasing temperature.
These results extend and generalize to finite temperature the results obtained in Ref.~\cite{SPS-2010} at zero temperature only.
A further comment on the physical relevance of the above behaviour will be made in Sec.~\ref{sec:Landau-criterion}.

\vspace{0.05cm}
\begin{center}
{\bf B. Temperature dependence \\ of the critical current}
\end{center}

\begin{figure}[t]
\begin{center}
\includegraphics[width=8.6cm,angle=0]{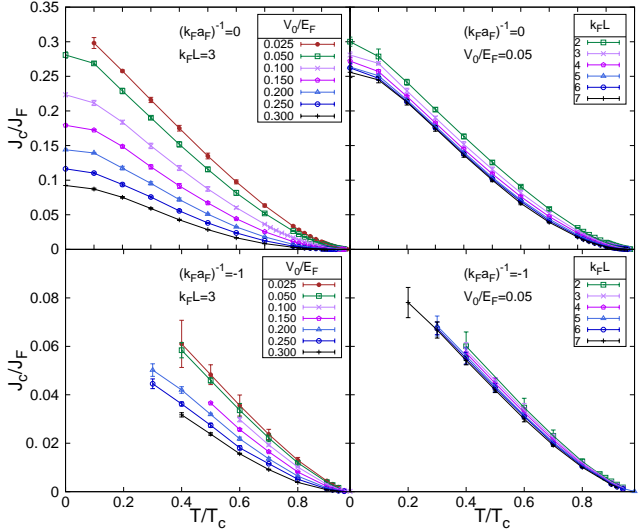}
\caption{(Color online): Temperature dependence of the critical current $J_{c}$ (in units of $J_{F}$), for a barrier of width $k_{F}L= 3$ and various heights (left panels), 
                                                                                               and a barrier of height $V_{0}/E_{F} = 0.05$ and various widths (right panels).
                                                                                               In each case, the upper and lower panels refer to couplings $(k_{F}a_{F})^{-1} = 0$ and $(k_{F}a_{F})^{-1} = -1$, respectively.}
\label{Figure-18}
\end{center}
\end{figure} 

Figure~\ref{Figure-18} shows the temperature dependence of the critical current $J_{c}$ for various barrier heights and widths and for two characteristic couplings.
Note that increasing the barrier height has a stronger effect on $J_{c}$ than increasing the barrier width.
Note also that, due to the intrinsic limitations of the LPDA approach as pointed out in Sec.~\ref{sec:theoretical_approach}-C, for the coupling  $(k_{F}a_{F})^{-1} = -1$ on the BCS side the temperature
dependence of $J_{c}$ can be determined with the exclusion of the temperature range $T \lesssim 0.3 \, T_{c}$.
For the limited set of barriers considered in Ref.~\cite{SPS-2010}, we have further verified that the values of the critical current obtained at unitarity by the LPDA approach in the limit of zero temperature compares favourably  
with those obtained in Ref.~\cite{SPS-2010} by solving the BdG equations at zero temperature only.

\vspace{0.05cm}
\begin{center}
{\bf C. Behaviour of the critical current \\ in the critical region}
\end{center}

\begin{figure}[t]
\begin{center}
\includegraphics[width=8.8cm,angle=0]{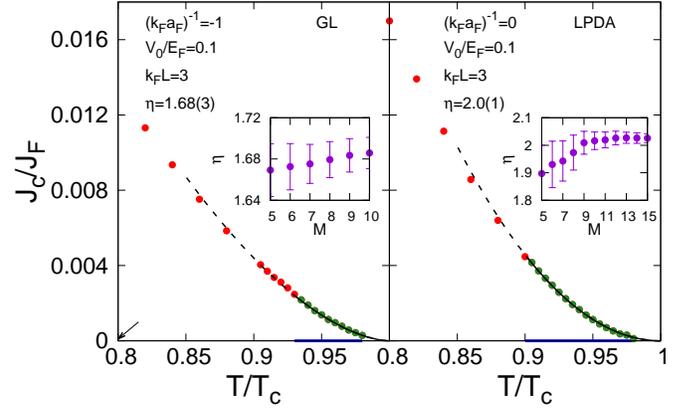}
\caption{(Color online): Numerical procedure to obtain the exponent $\eta$ for the vanishing of $J_{c}$ upon approaching $T_{c}$.
                                                                                              The same barrier is considered for two different couplings.
                                                                                              In both panels, the temperature interval of the fit (\ref{exponent-eta}) corresponds to blue thick segment, 
                                                                                              and the $M$ values of $J_{c}(T)/J_{F}$ over which the fit is applied correspond to the green dots (additional values of $J_{c}(T)/J_{F}$ are given by the red dots).
                                                                                              The black thick curve through the green dots gives the result of the fit, which is extrapolated outside its range by the dashed curve.
                                                                                              The oblique arrow in the left panel points to the minimum value of $J_{c}$ that can be handled numerically with confidence.
                                                                                              The insets reports the values of $\eta$ with the associated error bars, obtained by restricting the fit (\ref{exponent-eta}) to the first $M$ numerical values 
                                                                                              closer to the upper limit $T/T_{c}=0.98$ used in the fit.}
\label{Figure-19}
\end{center}
\end{figure} 

It is of interest to determine the behaviour of the critical current $J_{c}$ as it vanishes upon approaching the critical temperature $T_{c}$.
To this end, we associate to this behaviour of $J_{c}$ a positive exponent $\eta$, such that for $T$ sufficiently close to $T_{c}$
\begin{equation}
\frac{J_{c}(T)}{J_{F}} \simeq D \, t^{\eta}
\label{exponent-eta}
\end{equation}
\noindent
where $D$ is a numerical coefficient and $t = (T_{c}-T)/T_{c}$.
It turns out that the value of $\eta$ is not universal but depends on the type of junctions one considers for the Josephson effect \cite{Golubov-2004} (in our case, on the values of the barrier height and width for the SsS junction).
By our analysis, we will also find how $\eta$ depends on the coupling value $(k_{F}a_{F})^{-1}$.

For definiteness, we shall explicitly consider two coupling values, for which the analysis to extract the value of the exponent $\eta$ will be separately carried out with a considerable number of barriers in the following way:

\noindent
(i) For $(k_{F}a_{F})^{-1} = -1$ on the BCS side of the crossover, we find it convenient to use the results for $J_{c}(T)$ obtained by the GL equation, to which the results obtained by the LPDA equation have been shown in Sec.~\ref{sec:GL-regime}-B
to smoothly connect for $T$ is sufficiently close to $T_{c}$.
In particular, in this case we restrict $T/T_{c}$ in Eq.~(\ref{exponent-eta}) to the interval $0.93 \lesssim T/T_{c} \lesssim 0.98$, where $J_{c}(T)/J_{F}$ is calculated for up to $M=10$ distinct values of $T/T_{c}$.

\noindent
(ii) At the unitarity limit where $(k_{F}a_{F})^{-1} = 0$, $J_{c}(T)$ close to $T_{c}$ can be obtained directly by the LPDA equation.
In this case, $T/T_{c}$ in Eq.~(\ref{exponent-eta}) is restricted to the wider interval $0.90 \lesssim T/T_{c} \lesssim 0.98$, where $J_{c}(T)/J_{F}$ is calculated for up to $M=15$ distinct values of $T/T_{c}$.

In both cases, the choice of the upper value $T/T_{c}=0.98$ is related to the minimum value of $J_{c}(T)/J_{F}$ ($\simeq 1.0 \times 10^{-4}$), which can be handled with confidence by our numerical calculations for the shape of the
Josephson characteristics to be reliable.
The lower value $T/T_{c}=0.90$ considered at unitarity might be further reduced down to $T/T_{c}=0.85$ with no harm, to be contrasted with the lower value $T/T_{c}=0.93$ considered for $(k_{F}a_{F})^{-1} = -1$.
This difference is in line with the expectation about the width of the critical region near $T_{c}$ being larger at unitarity than in the BCS regime \cite{Lucheroni-2002,Taylor-2009}.
In the present context, this can be seen by casting the Ginzburg criterion for the relevance of fluctuations in the critical region \cite{Tilley-Tilley-1986} on the BCS side of unitarity, in the form
$t^{1/2} \! \left( \frac{k_{F} \xi_{\mathrm{pair}}}{2 \pi} \right)^{2} \lesssim 1$, where $\xi_\mathrm{pair}$ is the Cooper pair size at zero temperature \cite{PS-1994}, whose values can be read directly from 
Fig.~\ref{Figure-9} for the couplings of interest here.

Although our analysis to extract the exponent $\eta$ from the fit (\ref{exponent-eta}) could \emph{a priori} be influenced by the upper value of $T/T_{c}$ which can be reached numerically, in practice the $M$ numerical values 
of $J_{c}(T)/J_{F}$ on which the fit (\ref{exponent-eta}) is applied are distributed in a sufficiently smooth way, that the error on the exponent $\eta$ obtained within the restricted interval of $T/T_{c}$ we are considering does not exceed a few percent.
A typical example of this analysis is shown in Fig.~\ref{Figure-19} for the two couplings we are considering.
Here, the fit (\ref{exponent-eta}) is applied to the $M$ numerical values closer to $T_{c}$, ranging from $M=5$ to $M=10$ for coupling $(k_{F}a_{F})^{-1} = -1$ and from $M=5$ to $M=15$ for coupling $(k_{F}a_{F})^{-1} = 0$.
The values of the exponent $\eta$ obtained in this way for given value of $M$ (with the associated error bars) are reported in the insets of Fig.~\ref{Figure-19}.
The final estimates for $\eta$ (with the associated errors) given in the figure for the two couplings are eventually obtained by averaging over all values reported in the insets.

\begin{figure}[t]
\begin{center}
\includegraphics[width=8.8cm,angle=0]{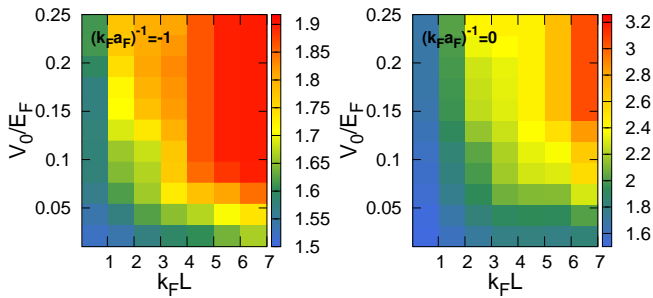}
\caption{(Color online): Colour maps of the exponent $\eta$ for the critical current $J_{c}$ in the critical region.
              Results are shown for two coupling values over extended ranges of the barrier height and width.}
\label{Figure-20}
\end{center}
\end{figure} 

We have extended the above analysis for both coupling values $(k_{F}a_{F})^{-1} = (-1,0)$ to a large set of barriers, to determine how the exponent $\eta$ evolves for increasing $V_{0}/E_{F}$ and $k_{F}L$.
The results of this extended analysis are given by the colour maps of Fig.~\ref{Figure-20}.
For the coupling $(k_{F}a_{F})^{-1} = -1$ on the BCS side of the crossover (left panel), the exponent $\eta$ is seen to evolve smoothly from the value $1.5$ (bottom-left corner) toward the value $2.0$ (upper-right corner),  
that correspond to ``high barrier transparency''  and ``low barrier transparency'', respectively \cite{Golubov-2004}. 
For the unitarity coupling $(k_{F}a_{F})^{-1} = 0$ (right panel), larger values of $\eta$ are instead systematically obtained upon approaching the upper-right corner of the corresponding colour map within the same ranges 
of $V_{0}/E_{F}$ and $k_{F}L$.
No one of the cases here considered appears to correspond to a point-contact constriction \cite{Golubov-2004} or a tunnel junction with small transmission probability \cite{A-B-1963}, for which $\eta=1$.

\section{Landau criterion at finite temperature with a vanishingly small barrier} 
\label{sec:Landau-criterion}

In this Section, we exploit the Josephson characteristics obtained by the LPDA approach at finite temperature, to determine how the Landau criterion for a superfluid Fermi gas evolves along the BCS-BEC crossover when the temperature is raised from $T=0$ up to the critical temperature $T_{c}$.

Quite generally, according to the Landau criterion for superfluidity \cite{Landau-1941}, the velocity of the superfluid flow should never exceed the corresponding value of the critical velocity past which the flow of the superfluid component becomes dissipative. 
This is because, for velocities of flow below this value, excitations cannot appear in the liquid which would destroy the superfluid flow \cite{footnote-Baym}. 
In the case of interest here, the critical velocity is determined by the onset of quasiparticle excitations which are of a different nature on the two sides of the BCS-BEC crossover, namely, pair-breaking excitations on the BCS side and sound-mode quanta on the BEC side. 
We will verify that the present calculation based on the LPDA approach captures this important physical feature on both sides of the crossover, because no stable numerical solution of the LPDA equation can consistently be found above the boundaries set by these two kinds of quasiparticle excitations (as represented by the dashed lines delimiting the shaded area in Fig.~\ref{Figure-22} below).

To this end, at any temperature $T$ in the interval $0 \le T \le T_{c}$ we adopt a similar procedure to that utilized at $T=0$ in Refs.~\cite{SPS-2007} and \cite{SPS-2010}, and for given coupling we determine the limiting value of $J_{c}$ for a vanishingly small barrier
which acts as an impurity probing the stability of the superfluid flow.
It will turn out that the limiting value of $J_{c}$ determined in this way decreases for increasing temperature at any coupling, as expected due to the emergence at finite temperature of a normal component 
with density $n_{\mathrm{n}} = n - n_{\mathrm{s}}$ \cite{AGD-1975} (cf. also Sec.~\ref{sec:theoretical_approach}-B).

\begin{figure}[t]
\begin{center}
\includegraphics[width=8.5cm,angle=0]{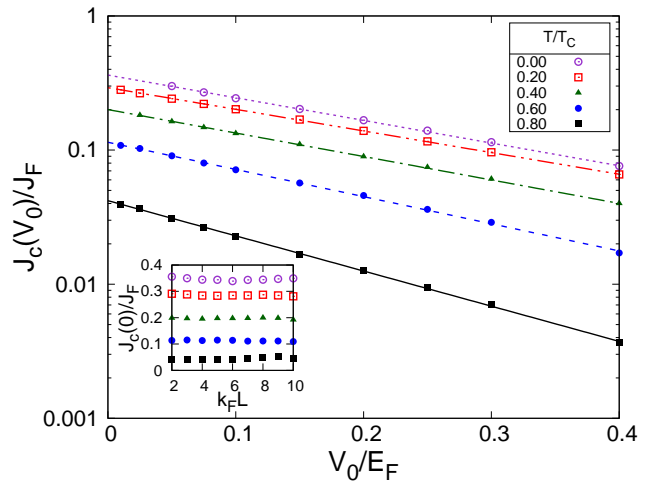}
\caption{(Color online): Semi-log plot of the critical current $J_{c}(V_{0})$ (in units of $J_{F}$) vs $V_{0}/E_{F}$ with a barrier of width $k_{F}L=2$,
              for $(k_{F}a_{F})^{-1}=0$ and various temperatures.
              The inset shows the behaviour of the limiting value of $J_{c}(0)=J_{c}(V_{0} \rightarrow 0^{+})$ when the width $L$ is varied, for the same coupling and temperatures of the main panel.}
\label{Figure-21}
\end{center}
\end{figure} 

For given coupling and temperature, we implement the limit of a vanishing barrier by considering progressively smaller values of the barrier height $V_{0}$ and width $L$ in the following way.
First, the width $L$ is fixed and the critical current $J_{c}$ is calculated for a set of decreasing values of the height $V_{0}$.
The values of $J_{c}(V_{0})$ obtained in this way are then fitted by the exponential form
\begin{equation}
\frac{J_{c}(V_{0})}{J_{F}} = E \, e^{-V_{0}/F}
\label{fitting-critical-current}
\end{equation}
where $E$ and $F$ are parameters. 
The required limiting value $J_{c}(0)=J_{c}(V_{0} \rightarrow 0^{+})$ is eventually obtained via graphical extrapolation.
This process is then repeated several times with progressively smaller values of the width $L$, in order to verify how does the value $J_{c}(0)$ obtained in this way evolve toward the limit $L \rightarrow 0^{+}$.
An example of this procedure is shown in Fig.~\ref{Figure-21} for $(k_{F}a_{F})^{-1}=0$ and several temperatures.
In this case, the range $0.01\!-\!0.4$ of $V_{0}/E_{F}$ proves sufficient to reproduce the exponential behaviour (\ref{fitting-critical-current}), thereby allowing the required extrapolation for $V_{0} \rightarrow 0^{+}$
as well as to verify the dependence of the results on $L$. 
We note that the occurrence of an exponential dependence of $J_{c}$ on $V_{0}$ for given $L$ may also be inferred from the data of Fig~13 in Ref.\cite{Zou-Dalfovo-2014}, 
although only at unitary and for $T=0$.

\begin{figure}[t]
\begin{center}
\includegraphics[width=8.7cm,angle=0]{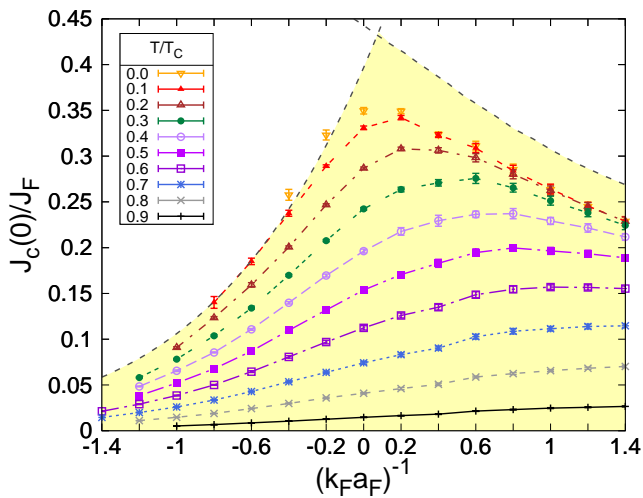}
\caption{(Color online): The critical current $J_{c}(0)$ (in units of $J_{F}$), for which the superfluid flow becomes unstable in the limit of a barrier 
              with vanishing height and width, is shown as a function of $(k_{F}a_{F})^{-1}$ for several temperatures.
              The left and right dashed lines (which cross each other near unitarity) correspond to the appearance at $T=0$ of pair-breaking and sound-mode excitations, respectively \cite{SPS-2010}.
              The shaded area highlights the region of allowed superfluid flow, where the superfluid critical current lies below both dashed lines.}
\label{Figure-22}
\end{center}
\end{figure} 

Through this procedure, a large set of values for the critical current $J_{c}(0)$ has been produced by analyzing about $30 000$ Josephson characteristics for different couplings, temperatures, and barrier heights and widths, in order to implement in each case the limit of a vanishing barrier.
The results of this systematic study are summarized in Fig.~\ref{Figure-22}, where the coupling dependence of the critical current $J_{c}(0)$ is shown for several temperatures in the
interval $0 \le T \le T_{c}$ \cite{footnote-Croitoru}.
These results implement the Landau criterion for superfluidity at finite temperature along the BCS-BEC crossover, which sets the boundary beyond which the superfluid flow becomes unstable in the limit of 
a vanishingly small barrier for the given temperature.
In this present context of a vanishing barrier, the analysis of Sec.~\ref{sec:critical-current}-C is consistent (within numerical errors) with the value $\eta = 1.5$ for the exponent of Eq.~(\ref{exponent-eta}) \emph{irrespective of coupling\/}, in line with the results of Fig.~\ref{Figure-20}.

Also shown in Fig.~\ref{Figure-22} are the boundary curves (dashed lines) delimiting the shaded region, which correspond to the occurrence at zero temperature of pair-breaking excitations on the BCS side of unitarity (left branch) and of sound-mode excitations on the BEC side of unitarity (right branch), respectively \cite{SPS-2010}.
Note that all numerical values of $J_{c}(0)$ obtained by the LPDA equation correctly fall within the shaded area for any temperature.
As already pointed out in a related context when solving the BdG equations at zero temperature \cite{SPS-2010}, it may appear rather remarkable that, \emph{in the presence of an infinitesimal seed that breaks translational symmetry\/},
the solutions of the LPDA equation here obtained also at finite temperature reflect the presence, not only of pair-breaking excitations characteristics of the BCS side of the crossover 
(which for a homogeneous system would be show up already at the mean-field level), but also of sound-mode excitations characteristics of the BEC side of the crossover (which for a homogeneous system would instead require 
the inclusion of pairing fluctuations beyond mean field).
This remark is in line with the discussion reported in Sec.~\ref{sec:spatial-profiles-out}-C while commenting on Fig.~\ref{Figure-9}.

\begin{figure}[t]
\begin{center}
\includegraphics[width=8.5cm,angle=0]{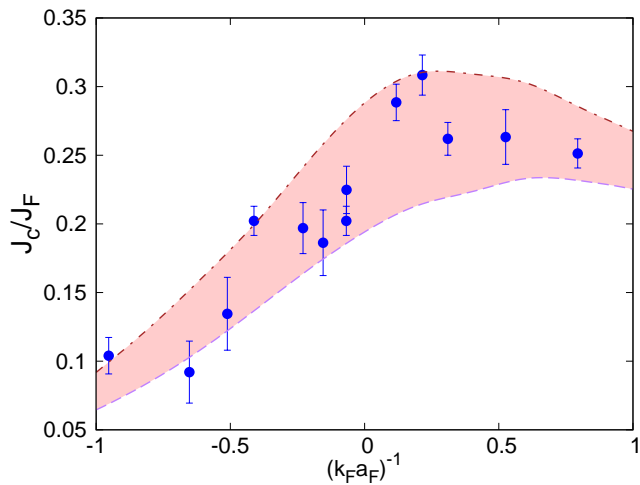}
\caption{(Color online): Comparison of the values of $J_{c}/J_{F}$ over an extended coupling range, obtained experimentally in Ref.~\cite{Moritz-2015} for a weak barrier (dots with error bars) and calculated theoretically for 
                                     a vanishingly small barrier in the temperature interval $0.2 \le T/T_{c} \le 0.4$ (shaded area).
                                     The upper and lower curves delimiting the shaded area correspond, respectively, to the values $T/T_{c} = 0.2$ and $T/T_{c} = 0.4$ in Fig.~\ref{Figure-22}.}
\label{Figure-23}
\end{center}
\end{figure} 

These apparently favourable outcomes, however, do not eliminate the need of a further inclusion of pairing fluctuations when dealing with inhomogeneous situations (like in the presence of even an infinitesimal barrier here considered), 
which are expected to quantitatively modify the results of Fig.~\ref{Figure-22} at finite temperature especially on the BEC side of the crossover \cite{NSR-1985}. 
This inclusion would not only result in more reliable values of $T_{c}$ relative to the Fermi temperature $T_{F}$ \cite{PPS-2018}, but should also specifically shift somewhat back toward unitarity the position of maximum of the curves in Fig.~\ref{Figure-22} at finite temperature. 
Implementing the inclusion of pairing fluctuations on top of inhomogeneous geometrical constraints would, however, require considerable theoretical efforts which are beyond the purposes of the present paper.

In any event, the theoretical results of Fig.~\ref{Figure-22} can be directly compared with the experimental data for the critical velocity of the superfluid flow obtained in Ref.~\cite{Moritz-2015} along the BCS-BEC crossover in the low-temperature regime.
This is because in that reference efforts were made to implement an experimental procedure that resembles as closely as possible the one envisioned theoretically by Landau, 
whereby an ultra-cold Fermi gas of $^{6}\mathrm{Li}$ atoms was perturbed by a ``weak'' (point-like) barrier moving through it with a steady velocity.
This comparison is shown in Fig.~\ref{Figure-23}, where the theoretical results from Fig.~\ref{Figure-22} within the temperature interval $0.2 \le T/T_{c} \le 0.4$ are represented by the shaded area.
With all the experimental data falling within the shaded area, one may conclude that a good agreement is obtained between experiment and theory under the present circumstances.

\section{Relationship between critical current and condensate density}
\label{sec:condensate-density}

Recently, a relationship has been put forward to connect the critical Josephson current with the condensate density \cite{Yang-1962} in the presence of a strong barrier, based on a theoretical argument originating from the BEC side of the crossover \cite{ZZ-2019}.
This relationship was then utilized to determine the condensate density from experimental measurements of the critical Josephson current, both in three \cite{Roati-2020} and in two  \cite{Moritz-2020} dimensions.
Here, we point out a different relationship between the critical Josephson current and the condensate density, which holds under somewhat complementary circumstances, namely, on the BCS side of the crossover for a weak barrier.
Specifically, our argument is based on the results of Sec.~\ref{sec:Landau-criterion} obtained with a vanishingly small barrier.

\begin{figure}[t]
\begin{center}
\includegraphics[width=8.7cm,angle=0]{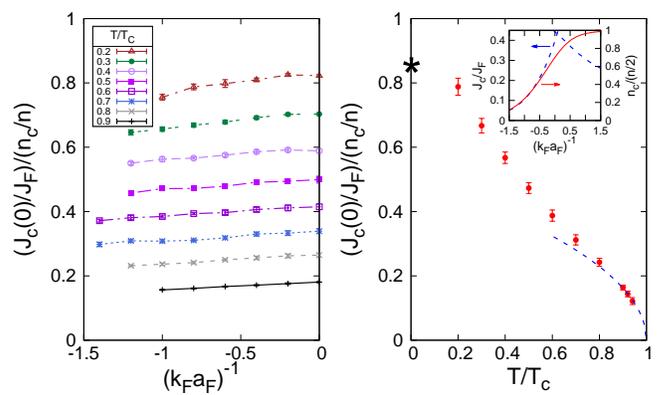}
\caption{(Color online): Left panel: Coupling dependence of the ratio between $J_{c}(0)/J_{F}$ taken from Fig.~\ref{Figure-22} and the condensate density $n_{c}$ 
                                      (in units of the bulk density $n$) obtained at the mean-field level \cite{Salasnich-2005} for various temperatures.
                                      Right panel: Temperature dependence of the ``universal'' function extracted from this ratio in the interval $0.2 \le T/T_{c} \le 0.9$.
                                       Here, the broken line corresponds to a square-root behaviour close to $T_{c}$ and the star for the value $8/(3 \pi)$ obtained analytically at $T=0$.
                                      The inset shows the coupling dependences of $J_{c}(0)/J_{F}$ (dashed line - left scale) and of $n_{c}/(n/2)$ (full line - right scale) at $T=0$, covering the whole BCS-BEC crossover.}
\label{Figure-24}
\end{center}
\end{figure} 

To this end, the left panel of Fig.~\ref{Figure-24} shows the ratio, between the values of $J_{c}(0)/J_{F}$ taken from Fig.~\ref{Figure-22} on the BCS side of unitarity from 
$(k_{F} a_{F})^{-1} = - 1.4$ to $(k_{F} a_{F})^{-1} = 0$ in the temperature interval $0.2 \le T/T_{c} \le 0.9$, and the condensate density $n_{c}$ (in units of the bulk density $n$) obtained at the mean-field level 
for the given coupling and temperature \cite{Salasnich-2005}.
Within numerical errors, one sees that, for given temperature, this ratio turns out to be essentially independent from coupling.
For each of the considered temperatures, we then determine the average value of this ratio over the whole coupling interval, with an associated error bar that corresponds to the spread between its maximum and minimum values.
The results of this procedure are represented by the ``universal'' function of temperature shown in the right panel of Fig.~\ref{Figure-24}, which vanishes at $T_{c}$ with a square-root behaviour as expected.
In addition, the inset in the right panel of Fig.~\ref{Figure-24} shows how the proportionality between $J_{c}(0)$ and $n_{c}$ holds on the BCS side of unitarity also at zero temperature, where analytic expressions of these quantities are available at the mean-field level \cite{SPS-2010} \cite{Salasnich-2005}.

\section{Comparison with experimental data for an ultra-cold Fermi gas with a finite barrier} 
\label{sec:comparison-experiment}

A further comparison between theory and experiment for $J_{c}$ at finite temperature, this time in the presence of a \emph{finite} barrier with well-defined parameters, can be obtained by comparing the results of the present LPDA approach
with the recent experimental data of Ref.~\cite{Roati-2020} taken also with an ultra-cold Fermi gas of $^{6}\mathrm{Li}$ atoms.

To make this comparison meaningful, care should be taken in converting the normalizations $J_{F}$ of the current, $E_{F}$ of the barrier height, and  $k_{F}^{-1}$ of the barrier width, from the trapped values used in the experiment to the homogeneous values used in the theoretical calculations.
To this end, what one needs to know is the ratio $\kappa = E_{F}^{t}/E_{F}^{0}$ between the \emph{trap\/} Fermi energy $E_{F}^{t} = \omega_{0} (3N)^{1/3}$ (where $\omega_{0}$ is the average trap frequency and $N$ the total atoms number) and the value $E_{F}^{0}=(k_{F}^{0})^{2}/(2m)$ associated with the local wave vector $k_{F}^{0}$ at the trap center, obtained by measuring the density profile of the atomic cloud.
This is because the moving barrier in the experiment of Ref.~\cite{Roati-2020} essentially explores the spatial region close to the trap center, thus implying that $k_{F}^{0}$ can be identified with the value of $k_{F}$ used in the theoretical calculations.
Accordingly, the values $V_{0}/E_{F}$ and $k_{F}L$ used in the theoretical calculations for the barrier height and width have to be identified, respectively, with $\kappa$ times the value $V_{0}/E_{F}^{t}$ and 
$\sqrt{\kappa}$ times the value $k_{F}^{t}L$ utilized in the experiment.
By the same token, the values of $J_{c}/J_{F}$ obtained by the theoretical calculations have to be identified with $\kappa^{-2}$ times the values $J_{c}/J_{F}^{t}$ obtained in the experiment at the corresponding couplings and temperatures.

\begin{figure}[t]
\begin{center}
\includegraphics[width=8.9cm,angle=0]{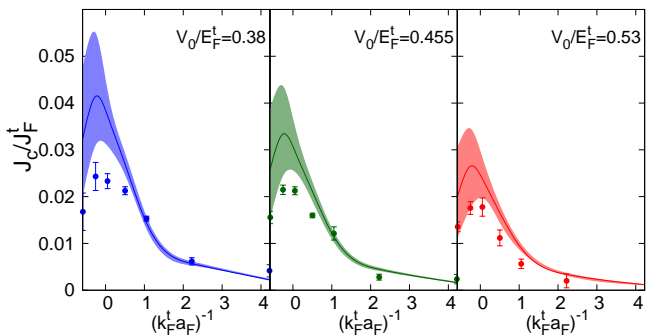}
\caption{(Color online): Critical current $J_{c}$ (in units of the trap value $J_{F}^{t}$) vs the experimental coupling parameter $(k_{F}^{t} a_{F})^{-1}$ for three different barrier heights
               (in all cases the barrier width is $k_{F}^{t} L = 2.530 \pm 0.125$).
               The experimental data from Ref.~\cite{Roati-2020} (dots with error bars) are compared with the theoretical values represented by the shaded areas, whose boundaries are set by the experimental uncertainties
               on $T/T_{F}^{t}$ and $V_{0}/E_{F}^{t}$ (see the text).
               For each experimental data point, the values of $(k_{F} a_{F})^{-1}$ and $T/T_{c}$ at which the theoretical calculations have been performed are listed in Table~\ref{Table-II}.}
\label{Figure-25}
\end{center}
\end{figure} 

In addition, to relate the value of the experimental temperature (in units of the trap Fermi temperature $T_{F}^{t}=E_{F}^{t}/k_{B}$) to that of the theoretical temperature (in units of $T_{c}$, as calculated within the present approach), it is sufficient to identify the theoretical value $T/T_{c}$ with the experimental value $T/T_{c}^{t} = (T/T_{F}^{t}) (T_{F}^{t}/T_{c}^{t})$.
Here, the values of the critical temperature $T_{c}^{t}/T_{F}^{t}$ in units of $T_{F}^{t}$ for a harmonic trap as a function of $(k_{F}^{t} a_{F})^{-1}$ can be taken from the fully self-consistent $t$-matrix calculation of Ref.~\cite{Pini-2020}
(cf. Fig.~7 therein), which is believed to well reproduce the thermodynamic properties of the attractive Fermi gas.
This kind of identification is common in condensed matter when comparing theoretical results with experimental data and has already been considered in the context of ultra-cold Fermi gases \cite{Sidorenkov-2013}.

Figure~\ref{Figure-25} shows the comparison between the theoretical values of the critical current obtained by the LPDA approach (shaded areas) and the corresponding experimental results 
(dots with error bars) taken from Fig.3 of Ref.~\cite{Roati-2020}, for three values of the barrier height and for several couplings across the BCS-BEC crossover.
All experimental values shown in the figure refer to a low-temperature regime with $T/T_{F}^{t}=0.06(2)$ \cite{Roati-2020}.
For convenience, the conversion from the experimental to the theoretical values of the coupling and the corresponding theoretical values of $T/T_{c}$ are listed in Table~\ref{Table-II} for all
the experimental data shown in Fig.~\ref{Figure-25}.

\begin{table}[h]
\begin{tabular}{c|c|c|c}
$(k_F^ta_F)^{-1}$ & $E_F^t/E_F$ & $(k_Fa_F)^{-1}$ & $T/T_c$\\
\hline
-0.583 & 0.795 & -0.52 & 0.344 \\
\hline
-0.254 & 0.712 & -0.214 & 0.233 \\
\hline
0.0516 & 0.577 & 0.039 & 0.163 \\
\hline
0.504 & 0.511 & 0.36 & 0.117 \\
\hline
1.055 & 0.386 & 0.655 & 0.076 \\
\hline
2.22 & 0.270 & 1.15 & 0.049 \\
\hline
4.22 & 0.206 & 1.91 & 0.034
\end{tabular}
\caption{Conversion from experimental to theoretical coupling values and corresponding theoretical values of $T/T_{c}$ for $T/T_{F}^{t}=0.06$, used for the comparison shown in Fig.~\ref{Figure-25} \cite{footnote-2}.}
\label{Table-II}
\end{table}

To comply with the experimental uncertainties on the values of $T/T_{F}^{t}$ and $k_{F}^{t} L$, in each panel of Fig.~\ref{Figure-25} the upper (lower) boundary of the shaded area associated with the theoretical calculations corresponds to $T/T_{F}^{t}=0.04$ ($T/T_{F}^{t}=0.08$), while at the same time the values of $k_{F}^{t} L$ are decreased (increased) by $5\%$ with respect to their nominal values (the full line within each shaded areas corresponds to the mean values of $T/T_{F}^{t}$ and $k_{F}^{t} L$).

Figure~\ref{Figure-25} evidences an overall good agreement between theory and experiment for a wide set of physical conditions.
In each case, the theoretical results appear to systematically overestimate the experimental values of the critical current, although the discrepancy decreases for increasing $V_{0}/E_{F}^{t}$.
This discrepancy may (at least partly) originate from the non-trivial geometrical shape of the experimental barrier transverse to the current flow, whose details are not taken into account in the present calculations 
but could affect the flow of the supercurrent and lead to a lost of coherence \cite{BHA-2020}. 
Although these kinds of geometrical effects could possibly be introduced within the LPDA approach, their consideration is beyond the scope of the present work.

\section{Concluding remarks}
\label{sec:conclusions}

In this paper, we have provided a detailed account of various aspects of the stationary Josephson effect at finite temperature, for a wide choice of barriers and an extended range of couplings along the BCS-BEC crossover.
To make these calculations feasible with reasonable efforts, we have resorted to solving the LPDA non-linear differential equation for the gap parameter, obtained by a coarse graining procedure on the BdG equations.
In this way, we have not only successfully reduced the computational time within acceptable limits, but especially much compressed the storage space that would otherwise be required when solving the BdG equations.
The numerical flexibility achieved in this way has allowed us to perform a systematic study of several aspects of the stationary Josephson effect for a wide choice of barriers in an SsS junction, not only at finite temperature 
but also by varying  the inter-particle attraction along the BCS-BEC crossover, from the BCS regime up to the unitarity limit and beyond.

The outcomes of this systematic study, obtained under a variety of circumstances for the Josephson characteristics (cf. Sec.~\ref{sec:Josephson-characteristics}) and the spatial profiles of the magnitude and phase of the gap parameter (cf. Sec.~\ref{sec:spatial-profiles-out}), that deserve a special mention are: 
The appearance of a novel kind of Josephson-induced proximity effect by increasing the barrier width (cf. Sec.~\ref{sec:spatial-profiles-in});
The gradual emergence of the GL regime out the LPDA approach for increasing temperature on the BCS side of unitarity (cf. Sec.~\ref{sec:GL-regime});
The smooth variation of the exponent for the critical current in the critical region upon varying the barrier height and width for given coupling (cf. Sec.~\ref{sec:critical-current});
The extension of the Landau criterion for superfluidity at finite temperature (cf. Sec.~\ref{sec:Landau-criterion});
The favourable comparison obtained with the experimental results currently available for ultra-cold Fermi gases (cf. Secs.~\ref{sec:Landau-criterion} and \ref{sec:comparison-experiment}). 

We have argued in more than one context (cf. Secs.~\ref{sec:spatial-profiles-out} and \ref{sec:Landau-criterion}) that, although the BdG approach (and therefore the associated LPDA approach) are based on 
a BCS mean-field decoupling \cite{BdG-1966}, their application to inhomogeneous situations (like the presence of a barrier in the Josephson effect) can lead to non-trivial outcomes, that would otherwise require consideration of pairing fluctuations in a homogeneous situation.
Nevertheless, it could be worthwhile to consider going beyond the LPDA approach and develop a double-coarse-graining procedure that would include pairing fluctuations over and above an inhomogeneous mean-field background,
possibly along the lines of Ref.~\cite{PPS-2004}.
This should be especially desirable in the light of possible technological applications based on superfluid Fermi systems far from the standard BCS behavior and embedded in complex geometrical constraints.


\begin{center}
\begin{small}
{\bf ACKNOWLEDGMENTS}
\end{small}
\end{center}

We are indebted to G. Deutscher for discussions and a critical reading of the manuscript, and to C. Di Castro and G. Roati for discussions.
Partial financial support from the Italian MIUR under Project PRIN2017 (20172H2SC4) is acknowledged.

\appendix   
\section{NUMERICAL PROCEDURES FOR SOLVING THE LPDA EQUATION IN THE PRESENCE OF A SUPERCURRENT}
\label{sec:numerical_procedures}

In this Appendix, a method is set up for the numerical solution of the LPDA equation (\ref{LPDA-equation}) in the presence of a supercurrent.
This method can as well be used to solve directly the GP and GL equations in their respective domains of validity, thus complementing other numerical methods already available for these cases \cite{Choi-2003,Milosevic-2010}.

For definiteness, we take the supercurrent directed along the $x$ axis and assume translational invariance along the $y$-$z$ plane. 
Accordingly, the problem effectively reduces to one dimension and Eq.~(\ref{LPDA-equation}) with $A(x) \rightarrow - Q_{0}$ simplifies as follows:
\begin{eqnarray}
-\dfrac{m}{4\pi a_F}\tilde{\Delta}(x) & = &\mathcal{I}_0(x)\tilde{\Delta}(x)+\dfrac{\mathcal{I}_1(x)}{4m}\dfrac{d^2}{dx^2}\tilde{\Delta}(x)
\nonumber \\
& + & i \mathcal{I}_1(x) \dfrac{Q_0}{m} \frac{d \tilde{\Delta}(x)}{dx}
\label{eq:A_1}
\end{eqnarray}
where $\tilde{\Delta}(x)=e^{-2iQ_{0}x}\Delta(x) \equiv |\tilde{\Delta}|(x) \, e^{2 i \phi(x)}$ and
the coefficients $\mathcal{I}_0(x)$ and $\mathcal{I}_1(x)$ are given by the expressions (\ref{I_0}) and (\ref{I_1}) of the main text, respectively.
Through the integrals over the wave vector $\mathbf{k}$ in these coefficients, the system preserves memory of the orthogonal dimensions $y$ and $z$.
In addition, for a symmetric barrier with respect to $x=0$, the domain of solution of Eq.~(\ref{eq:A_1}) can be restricted to $(0,+\infty)$. 

It is further convenient to separate the real and imaginary parts of Eq.~(\ref{eq:A_1}) and introduce the spatial derivatives of $|\tilde{\Delta}|(x)$ and $\phi(x)$.
In this way, Eq.~(\ref{eq:A_1}) reduces to a system of four first-order differential equations in the unknown functions $|\tilde{\Delta}|(x)$, $|\tilde{\Delta}|'(x)$, $\phi(x)$, and $\phi'(x)$:
\begin{eqnarray}
|\tilde{\Delta}|' & = & \dfrac{d|\tilde{\Delta}|}{dx}
\label{eq:A_2-a} \\
\phi' & = & \dfrac{d\phi}{dx}
\label{eq:A_2-b} \\
\dfrac{d|\tilde{\Delta}|'}{dx} & = & 4 |\tilde{\Delta}| \! \left[ \phi'^{2} \! + \! 2Q_{0}\phi' \! - \! \frac{m}{\mathcal{I}_{1}} \! \left( \! \dfrac{m}{4\pi a_F}+\mathcal{I}_{0} \! \right) \! \right] 
\label{eq:A_2-c} \\
\dfrac{d\phi'}{dx} & = & - 2 \, \frac{|\tilde{\Delta}|'}{|\tilde{\Delta}|} \,\left( Q_{0} + \phi' \right) 
\label{eq:A_2-d}
\end{eqnarray}
\noindent 
where the $x$ dependence of the various quantities has been omitted for simplicity.
To ensure the solutions to these equations to be physically meaningful, we adopt the following \emph{boundary conditions}: 
\begin{eqnarray}
|\tilde{\Delta}|'(x=0) & = & 0 \, , \hspace{1.6cm} \phi(x=0)=0 \, , 
\nonumber \\
\lim_{x\rightarrow +\infty} |\tilde{\Delta}(x)| & = & \Delta_{0} \, ,  \hspace{0.2cm} \lim_{x\rightarrow+\infty}\phi'(x) = 0 \, , 
\label{eq:A_3}
\end{eqnarray}
where $\Delta_{0}$ is the bulk value of the gap parameter in the presence of the current.

We remark that Eq.~(\ref{eq:A_2-d}) entails the local current conservation in both limits of weak coupling and $T \rightarrow T_{c}^{-}$ \cite{Jacobson-1965} (when the LPDA equation recovers the GL equation) 
and of strong coupling and $T=0$ \cite{Hakim-1997} (when the LPDA equation recovers the GP equation).
However, when the BCS-BEC crossover is spanned at arbitrary temperature in the superfluid phase like in the present context, Eq.~(\ref{eq:A_2-d}) does not automatically guarantee current conservation.
For this reason, we have followed the strategy adopted in Ref.~\cite{SPS-2010}, where the Josephson problem was considered at $T=0$ throughout the BCS-BEC crossover, 
and replaced the imaginary part (\ref{eq:A_2-d}) of the gap equation by the condition (\ref{local-condition}) for local current conservation.
With the LPDA expression (\ref{two-fluid-current}) for the current, this local condition reads: 
\begin{eqnarray}
0 & = & j(x) -J = \frac{\phi'(x)}{m} \, n(x) + \frac{Q_{0}}{m} \left( n(x) - n \right) 
\label{eq:A_4} \\
& + & 2 \! \int \!\!\! \frac{d \mathbf{k}}{(2\pi)^{3}} \frac{\mathbf{k}}{m} \! \left[ f_{F}(E_{+}^{\mathbf{Q}_{0}}(\mathbf{k}|x)) - f_{F}(E_{+}^{\mathbf{Q}_{0}}(\mathbf{k}|x\rightarrow +\infty)) \right]
\nonumber
\end{eqnarray}
where $\mathbf{Q}_{0} = Q_{0} \hat{x}$, $n(x)$ is the value of the local density given by Eq.~(\ref{local-density}), and $n$ is the corresponding bulk density far away from the barrier.
The expression (\ref{eq:A_4}) depends on $|\tilde{\Delta}|(x)$ and $\phi'(x)$, and can be brought to the form of Eqs.~(\ref{eq:A_2-a})-(\ref{eq:A_2-c}) by moving $\phi'(x)$ to the left-hand side.

Finally, as it was done in Ref.~\cite{SPS-2010}, the wave vector $Q_{0}$ occurring in Eqs.~(\ref{eq:A_2-a})-(\ref{eq:A_2-c}) and (\ref{eq:A_4}) can also be considered as an independent variable, 
thereby imposing the additional boundary condition 
\begin{equation}
\phi(+ \infty) = \int_{0}^{+\infty} \!\!\! dx \, \phi'(x) = \frac{\delta\phi}{2}
\label{asymptotic-phase-difference-A}
\end{equation}
that fixes in advance the asymptotic phase difference (\ref{phase-difference}).

\vspace{0.05cm}
\begin{center}
{\bf 1. Implementing the implicit Runge-Kutta method}
\end{center}

The four equations (\ref{eq:A_2-a})-(\ref{eq:A_2-c}) and (\ref{eq:A_4}) to be solved are highly nonlinear in the variables $|\tilde{\Delta}|(x)$, $\phi'(x)$, and $Q_{0}$.
We have thus found it convenient to solve them through an implicit rather than an explicit Runge-Kutta method, since the latter might be subject to numerical instabilities \cite{SM-2003}.
Accordingly, we have implemented this method in the following way.

Owing to the spatial localization of the perturbance introduced by the barrier in an otherwise homogeneous superfluid, it is sufficient to restrict the solution of the equations 
(\ref{eq:A_2-a})-(\ref{eq:A_2-c}) and (\ref{eq:A_4}) to a finite interval extending from $x=0$ to $x=x_{\mathrm{max}}$ (in practice, $x_{\mathrm{max}}$ ranges from $10 k_{F}^{-1}$ to $200 k_{F}^{-1}$, 
depending on coupling and temperature).
The interval $(0,x_{\mathrm{max}})$ is then split into $N$ subintervals (not necessarily of equal length) identified by an index $\nu = (1,\cdots,N)$, such that $X_{\nu=0}=0$ and $X_{\nu=N}=x_{\mathrm{max}}$.
In each of these subintervals a mesh of $K$ points 
\begin{equation}
x_{k} = X_{\nu-1}+u_{k} \left( X_{\nu} - X_{\nu-1} \right) \hspace{0.5cm} (k = 1,\cdots,K)
\label{eq:A_5}
\end{equation}
is further selected such that $X_{\nu-1} \le x_{k} \le X_{\nu}$, with the condition $u_{k=1}=0$ and $u_{k=K}=1$ on the variables $u_{k}$.
In this way, each interval $(X_{\nu-1},X_{\nu})$ of variable length is mapped onto the interval $(0,1)$ of unit length.
[In practice, the values $N=30$ and $K=5$ have proven sufficient for good numerical convergence of the method.]

With this geometrical setting, the solution of Eqs.~(\ref{eq:A_2-a})-(\ref{eq:A_2-c}) and (\ref{eq:A_4}) proceeds as follows.
Within the $\nu$-th subinterval, these equations have the form
\begin{equation}
\mathbf{y'}(x_{k})=\mathbf{g}(\mathbf{y}(x_{k}),x_{k}) 
\label{eq:A_6}
\end{equation}
where $\{x_{k};k = (1,\cdots,K)\}$ is the mesh of points (\ref{eq:A_5}) and $\mathbf{y}^T=(\tilde{|\Delta|},\phi,\tilde{|\Delta|}',\phi')$ refers to the four unknown functions to be determined.
The left-hand side of Eq.~(\ref{eq:A_6}) is then represented by the expression
\begin{equation}
\mathbf{y}'(x_{k})=\sum_{i=1}^{K} \mathbf{q}_{i} \, f_{i}(x_{k})
\label{eq:A_7_1}
\end{equation}
in terms of $K$ distinct functions $\{f_{i}(x);i=(1,\cdots,K)\}$, one of which assumes the unit value in correspondence to one of the points $\{x_{k}\}$, that is to say
\begin{equation}
f_{i}(x_{k})=\delta_{ik} \, .
\label{eq:A_7_2}
\end{equation}
The coefficients $\{\mathbf{q}_{i};i = (1,\cdots,K)\}$ of the expansion (\ref{eq:A_7_1}), in turn, correspond to the values of the right-hand side of Eq.~(\ref{eq:A_6}), since
\begin{equation}
\mathbf{g}(\mathbf{y}(x_{k}),x_{k}) = \sum_{i=1}^{K} \mathbf{q}_{i} \, f_i(x_{k}) = \mathbf{q}_{k} 
\label{eq:A_8}
\end{equation}
for any given $k$.
Within $\nu$-th subinterval, we thus write:
\begin{eqnarray}
\mathbf{y}(x_{k}) & = & \mathbf{y}(X_{\nu-1}) + \sum_{i=1}^{K} \mathbf{q}_{i} \! \int_{X_{\nu-1}}^{x_k} \!\!\! dx \, f_{i}(x) 
\nonumber \\
& = & \mathbf{y}(X_{\nu-1}) + \sum_{i=1}^{K} a_{ki } \, \mathbf{q}_{i}
\label{eq:A_9_1}
\end{eqnarray}
with the short-hand notation 
\begin{equation}
a_{ki} = \int_{X_{\nu-1}}^{x_{k}} \!\!\! dx \, f_{i}(x) \, .
\label{eq:A_9_2}
\end{equation}
In particular, for $k=K$ such that $x_{K} = X_{\nu}$, Eq.~(\ref{eq:A_9_1}) becomes
\begin{equation}
\mathbf{y}(X_{\nu}) = \mathbf{y}(X_{\nu-1}) + \sum_{i=1}^{K} b_{i} \, \mathbf{q}_{i} 
\label{eq:A_9_3}
\end{equation}
where now
\begin{equation}
b_{i} = \int_{X_{\nu-1}}^{X_{\nu}} \!\!\! dx \, f_{i}(x) \, .
\label{eq:A_9_4}
\end{equation}
The result (\ref{eq:A_9_3}) is interpreted as yielding the solution $\mathbf{y}_{\nu}$ at the left side of the $\nu$-th subinterval in terms of the solution
$\mathbf{y}_{\nu-1}$ at the left side of the $(\nu-1)$-th subinterval.
Formally, this transfer of information from the $(\nu - 1)$-th to the $\nu$-th intervals can be cast in the form $\mathbf{y}_{\nu} = \mathcal{P}(\mathbf{y}_{\nu-1})$ in term of a ``propagator'' $\mathcal{P}$.

A convenient choice for the set of variables $\{u_{k}\}$ and for the functions (\ref{eq:A_7_2}) is obtained in terms of the normalized Legendre polynomials 
$\overline{P}_{n}(u) = \sqrt{\frac{2n+1}{2}} P_{n}(u)$ where $P_{n}(u)$ are standard Legendre polynomials \cite{MOS-1966}, such that
\begin{equation}
\int_{-1}^{+1} \!\!\! du \, \overline{P}_{n}(u) \, \overline{P}_{n'}(u) = \delta_{nn'} \, .
\label{eq:A_12_1}
\end{equation}
Let $\{\overline{u}_{k};k=(1,\cdots,K)\}$ be the $K$ distinct real zeros of the Legendre polynomial $P_{K}(u)$, obtained by solving the $K \times K$ eigenvalue problem that
results by cyclic application of the recurrence relation $(n+1) P_{n+1}(u) = (2n+1) u P_{n}(u) - n P_{n-1}(u)$ from $n=0$ up to $n=K-1$ where $P_{-1}(u)=0$
(cf., e.g., Appendix B of Ref.~\cite{SS-2017}).
From the normalization of the corresponding eigenvectors one also obtains the weight factors $\{w_{k};k=(1,\cdots,K)\}$ that enter the Gaussian quadrature:
\begin{equation}
\int_{-1}^{+1} \!\!\! du \, \overline{P}_{n}(u) \, \overline{P}_{n'}(u) = \sum_{k=1}^{K}\overline{P}_{n}(\overline{u}_{k})\overline{P}_{n'}(\overline{u}_{k}) \, w_{k} = \delta_{nn'} \, .
\label{eq:A_12_2}
\end{equation}
In this way, one defines the orthogonal $(K \times K)$ matrix
\begin{equation}
S_{nk} = \overline{P}_{n}(\overline{u}_{k}) \, \sqrt{w_{k}} 
\label{eq:A_12_3}
\end{equation}
such that
\begin{equation}
\sum_{k=1}^{K} \, S_{nk} \, S^{T}_{kn'} = \delta_{nn'} \hspace{0.3cm} \mathrm{and} \hspace{0.3cm} \sum_{n=0}^{K-1} \, S^{T}_{kn} \, S_{nk'} = \delta_{kk'} \, .
\label{eq:A_12_4}
\end{equation}
Owing to Eq.~(\ref{eq:A_12_3}) and the second of Eqs.~(\ref{eq:A_12_4}), the requirement (\ref{eq:A_7_2}) is thus implemented in the form
\begin{eqnarray}
f_{i}(x_{k}) & = & \sqrt{w_{i}} \, \sum_{n=0}^{K-1} \, S^{T}_{in} \, \overline{P}_{n}(\overline{u}_{k})  
\label{special-chioce} \\
& = & \sqrt{w_{i}} \, \sum_{n=0}^{K-1} \, S^{T}_{in} \, \overline{P}_{n} \!\! \left( \! \frac{2 x_{k} - X_{\nu} - X_{\nu-1}}{h_{\nu}} \! \right)
\nonumber
\end{eqnarray}
where $h_{\nu} = X_{\nu} - X_{\nu-1}$ is the length of the $\nu$-th interval.
Here, the choice $2 u_{k} = \overline{u}_{k} + 1$ maps the $K$ roots $\overline{u}_{k}$ of the polynomial $P_{K}(u)$
in the interval $(-1,+1)$ into the $K$ values $u_{k}$ that enter Eq.~(\ref{eq:A_5}) and are restricted to the interval $(0,1)$.
In this way, the coefficients (\ref{eq:A_9_2}) become:
\begin{equation}
a_{ki} = \frac{h_{\nu}}{2} \, \sqrt{w_{i}} \, \sum_{n=0}^{K-1} \, S^{T}_{in} \, \int_{-1}^{\overline{u}_{k}} \!\! dy \, \overline{P}_{n}(y)
\label{final-expression-a}
\end{equation}
where
\begin{eqnarray}
& & \int_{-1}^{\overline{u}_{k}} \!\! dy \, \overline{P}_{n}(y) = \frac{1}{\sqrt{2n+1}} 
\label{integral-identity-Legendre-polynomials} \\
& \times & \left[ \frac{\overline{P}_{n+1}(\overline{u}_{k}) - \overline{P}_{n+1}(-1)}{\sqrt{2n+3}} - 
\frac{\overline{P}_{n-1}(\overline{u}_{k}) - \overline{P}_{n-1}(-1)}{\sqrt{2n -1}} \right] 
\nonumber
\end{eqnarray}
which results from the identity 
\begin{equation}
(2n+1) P_{n}(x) = \frac{d P_{n+1}(x)}{dx} - \frac{d P_{n-1}(x)}{dx}
\label{differential-identity-Legendre-polynomials}
\end{equation}
among the Legendre polynomials \cite{MOS-1966}.
Similarly, an expression like (\ref{final-expression-a}) holds for the coefficients (\ref{eq:A_9_4}), where the upper end of the integral in Eq.(\ref{integral-identity-Legendre-polynomials}) is now $+1$.

\vspace{0.05cm}
\begin{center}
{\bf 2. Successive steps of the Newton method}
\end{center}

There remains to apply a multi-dimensional Newton method for solving Eq.(\ref{eq:A_8}), within each of the $N$ subintervals in which the domain $(0,x_{\mathrm{max}})$ of the solution of this equation has been partitioned, as well as to solve Eq.~(\ref{eq:A_9_3}) that connects adjacent subintervals.
Accordingly, two successive steps have been set up for achieving the self-consistency of the solution.

\vspace{0.1cm}
\noindent
\emph{First step -}
With the help of Eqs.~(\ref{eq:A_5}) and (\ref{eq:A_9_1}), Eq.(\ref{eq:A_8}) is cast in the form
\begin{equation}
\mathbf{g} \! \left( \! \mathbf{y}(X_{\nu-1}) + \sum_{i=1}^{K} a_{ki } \, \mathbf{q}_{i}, X_{\nu-1} + h_{\nu} u_{k} \! \right) \! - \mathbf{q}_{k} = 0 
\label{eq:A_15_1}
\end{equation}
where $\nu = (1,\cdots,N)$ identifies the $\nu$-th subinterval and $k=(1,\cdots,K)$ the mesh of points in each subinterval.
For any given $\nu$ in Eq.~(\ref{eq:A_15_1}), the $K$ vectors $\{\mathbf{q}_{i}\}$ (each of $4$ dimensions) are considered unknown, while the values of $\mathbf{y}(X_{\nu-1})$ 
as well as the value of $Q_{0}$ are guessed beforehand as inputs of the calculation. 
One cycle of Newton method is then applied to determine the values of $\{\mathbf{q}_{i}\}$.

\vspace{0.1cm}
\noindent
\emph{Second step -}
Once the quantities $\{\mathbf{q}_{i}\}$ have been determined in this way for any given $\nu$, their values are inserted into Eq.~(\ref{eq:A_9_3}) which can be rewritten in the compact form 
\begin{equation}
\mathbf{y}_{\nu} = \mathcal{P}(\mathbf{y}_{\nu-1}) \hspace{0.3cm} \mathrm{with} \hspace{0.3cm} \nu = (1,\cdots,N) \, .
\label{eq:A_15_2}
\end{equation} 
These $4N$ equations contain $(N+1)$ sets of $4$-dimensional unknown quantities $\mathbf{y}_{\nu}$, since $\mathbf{y}_{\nu=0}$ at the left edge of the interval $(0,x_{\mathrm{max}})$
has also to be taken into account.
Including in $\mathbf{y}_{\nu=0}$ also the value of $Q_{0}$ that has to be self-consistently determined,
the required additional equations are supplied by the five boundary conditions (\ref{eq:A_3}) and (\ref{asymptotic-phase-difference-A}).
One cycle of Newton method is then applied to determine the values of $\{\mathbf{y}_{\nu};\nu=(0,\cdots,N)$.

\vspace{0.1cm} 
\noindent
\emph{Cycling back and forth -}
Once the quantities $\{\mathbf{y}_{\nu};\nu = (0,\cdots,N)\}$ have been determined in this way, their values are fed back into the first step above and new values for the quantities 
$\{\mathbf{q}_{i}\}$ are determined from Eq.~(\ref{eq:A_15_1}) through a second cycle of Newton method.
These values of $\{\mathbf{q}_{i}\}$ are then fed into the second step above, and new values of $\{\mathbf{y}_{\nu}\}$ (including $Q_{0}$) are in turn obtained.
Typically $5-30$ (depending on coupling and temperature) cycles of this two-step Newton method are required to eventually achieve full self-consistency.

In practice, the initial values of $\{\mathbf{y}^{T}_{\nu}\}$ are taken to be $(\tilde{|\Delta|}=\tilde{\Delta}_{0},\phi=0,\tilde{|\Delta|}'=0,\phi'=0)$ irrespective of $\nu$, as well as $Q_{0}=0$,
where $\tilde{\Delta}_{0}$ is the value of the gap parameter in the absence of the barrier.
In this way, the value $\delta\phi=0$ is initially assumed for the boundary condition (\ref{asymptotic-phase-difference-A}).
Cycling back and forth between the first and second steps above produces eventually a nontrivial spatial profile for $\tilde{|\Delta|}(x)$, although still with $\phi(x)=0$.
This profile is then taken as input for determining the next value of the Josephson characteristic for a non-vanishing value of $\delta\phi$ 
and the corresponding value of $Q_{0}$.
The process is repeated several times for increasing values of $\delta\phi$.

\vspace{0.05cm}
\begin{center}
{\bf 3. Two types of cycles of the Newton method}
\end{center} 

The numerical method just described is computationally quite demanding, to the extent that it becomes progressively more difficult to reach convergence for increasing values 
of $\delta\phi$.
This is especially true when considering the decreasing branch of the Josephson characteristics in order to approach the limiting value $\delta\phi = \pi$. 

Under these circumstances, we found it more convenient to alternate two types of cycles while searching for the solution, namely, 
a first type of cycle when $\delta\phi$ is fixed and $Q_{0}$ is consistently determined as described above,
and a second type of cycle when $Q_{0}$ is instead fixed and $\delta\phi$ is consistently determined.
In the latter case, the boundary condition (\ref{asymptotic-phase-difference-A}) is removed from the second step above and $\delta\phi$ is obtained by integrating the spatial profile of $\phi'(x)$.
At the same time, $Q_{0}$ is fixed at the value previously obtained by performing the first type of cycle.
The new solution, attained in this way by performing the second type of cycle, is then used back as input for the first type of cycle.
Typically, repeating this procedure a couple of times proves sufficient. 

\vspace{0.05cm}
\begin{center}
{\bf 4. Further numerical insights}
\end{center}

The LPDA equation (\ref{eq:A_1}) contains the local coefficients $\mathcal{I}_0(x)$ and $\mathcal{I}_1(x)$, which have to be determined from the expressions (\ref{I_0}) and (\ref{I_1}) 
with sufficient numerical accuracy at each step of the Newton method.
Comparable accuracy is also required to determine the local number density $n(x)$ and current $j(x)$ that enter the condition (\ref{eq:A_4}).
This is because the presence of numerical noise may render the convergence of the whole method quite difficult and sometimes even impossible to reach.
For these reasons, before starting the convergence procedure described above, all these quantities have been accurately evaluated over a regular grid of points in the variables
$(|\Delta|,\mu,Q_{0})$, and a tri-linear interpolation has been utilized to determine the actual values of these quantities needed at each iteration of the Newton method.

In practice, for a Josephson characteristics with a mesh of $50$ values of $\delta\phi$, the computational time ranges from $100$ to $800$ minutes, $15$ minutes of which are used for the calculation of the coefficients 
$\mathcal{I}_0(x)$ and $\mathcal{I}_1(x)$ of the LPDA equation over the above three-dimensional grid $(|\Delta|,\mu,Q_{0})$. 
This wide range of computational time reflects the convergence of a single profile $\Delta(x)$ for given $\delta\phi$, which may require from a few seconds up to $20$ minutes depending on the settings. 
The computer code (with no parallelization) was run on a devise with $3000$ Mhz CPU speed, $187$ GB RAM memory, $15$ GB swap memory, $1$ GB/s writing speed, and $250$ MB/s reading speed. 

\newpage
	


\begin{thebibliography}{99}
\bibitem{footnote-1} In Ref.~\cite{SS-2014}, the expression for the local superfluid density was further expanded at the lowest significant order in $\nabla \varphi(\mathbf{r})/2$.
                                Here, we avoid this expansion because, in practice, we shall deal with numerical values of the supercurrent that exceed this lowest-order approximation.
\end{thebibliography}
\end{document}